\documentclass[aps,prd,eqsecnum,twocolumn,showpacs,amssymb,nofootinbib,superscriptaddress]
{revtex4-1}

\usepackage{amsmath,amsthm,amsfonts}
\usepackage{hyperref}

\begin{document}

\title{Inhomogeneous Loop Quantum Cosmology: Hybrid Quantization of the
Gowdy Model}
\author{L. J. Garay}\email{luis.garay@fis.ucm.es}
\affiliation{Departamento de F\'{i}sica Te\'{o}rica II, Universidad
Complutense de Madrid, 28040 Madrid, Spain}
\affiliation{Instituto de Estructura
de la Materia, CSIC, Serrano 121, 28006 Madrid, Spain}
\author{M. Mart\'{i}n-Benito}
\email{merce.martin@iem.cfmac.csic.es} \affiliation{Instituto de
Estructura de la Materia, CSIC, Serrano 121, 28006 Madrid, Spain}
\author{G. A. Mena Marug\'{a}n}\email{mena@iem.cfmac.csic.es}
\affiliation{Instituto de Estructura de la Materia, CSIC, Serrano
121, 28006 Madrid, Spain}

\begin{abstract}

The Gowdy cosmologies provide a suitable arena to
further develop Loop Quantum Cosmology, allowing the
presence of inhomogeneities. For the particular case
of Gowdy spacetimes with the spatial topology of a
three-torus and a content of linearly polarized
gravitational waves, we detail a hybrid quantum theory
in which we combine a loop quantization of the
degrees of freedom that parametrize the subfamily of
homogeneous solutions, which represent Bianchi I
spacetimes, and a Fock quantization of the
inhomogeneities. Two different theories are
constructed and compared, corresponding to two
different schemes for the quantization of the Bianchi
I model within the {\sl improved dynamics} formalism
of Loop Quantum Cosmology. One of these schemes has
been recently put forward by Ashtekar and
Wilson-Ewing. We address several issues including the
quantum resolution of the cosmological singularity,
the structure of the superselection sectors in the
quantum system, or the construction of the Hilbert
space of physical states.
\end{abstract}


\maketitle

\section{Introduction}
\label{sec:intro}

In the absence of a full theory of canonical quantum
gravity, it is most instructive to study
symmetry-reduced systems in order to test and develop
mathematical techniques that can help in achieving the
quantization of general relativity, as well as to
progress in the understanding of the physical
phenomena emerging in quantum gravity. The symmetry
reduction makes these systems more manageable for a
complete quantization and facilitates the extraction
of physical predictions. In particular, mimicking the
techniques applied in Loop Quantum Gravity (LQG)
\cite{lqg}, a quantization program for homogeneous
models, known as Loop Quantum Cosmology (LQC)
\cite{lqc}, has been developed in recent years (see
e.g.
\cite{abl,aps3,mmo,iso1,iso2,iso3,chio,chi2,mmp,awe1,awe2}).
The so-called {\sl polymeric} quantization applied to
these systems has interesting physical consequences.
Remarkably, the analog of the classical cosmological
singularity is not present in the quantum theory and is
therefore resolved.

In view of the success of LQC, it is compelling to
extend the application of its techniques to the
quantization of inhomogeneous cosmological systems.
Several reasons motivate this extension. On the one
hand, this kind of models preserves one of the most
characteristic features of the full theory, namely the
field-like nature of the degrees of freedom. Thus, one
would expect that their quantization gives trustable
insights about some of the open questions in LQG which
are related to the presence of an infinite number of
degrees of freedom. On the other hand, it is essential
to investigate the role played by inhomogeneities in
the quantum theory in order to understand the
physical laws that explain the origin and evolution of
the Universe, and reach in this way realistic
predictions that can be confronted with cosmological
observations. Furthermore, the analysis of the
inhomogeneities would allow us to check the robustness
of the results obtained in homogeneous LQC, in
particular those concerning the quantum resolution of
classical cosmological singularities.

With this aim, we have carried out a complete
quantization of the linearly polarized Gowdy $T^3$
model \cite{gowd}. We have chosen this model because
it is the simplest cosmological system in vacuo which
contains inhomogeneities. Indeed, not only its classical
solutions are well known \cite{mon,ise}, but
also their quantization (by more conventional methods)
has deserved much attention since the 70s
\cite{qGow}. Actually, a complete Fock quantization of
the deparametrized system has been achieved
\cite{men1}, which has been shown to be essentially
unique \cite{men3}. Unlike these previous analyses,
which involve standard non-polymeric techniques, we
will discuss here a hybrid quantization that
combines the polymeric procedures of LQC ---applied to
the homogeneous sector of the system, namely the set
of degrees of freedom that describe the subfamily of
homogeneous solutions of the Gowdy model--- with the
Fock quantization of the inhomogeneities. Since the
full polymeric quantization of the model is an
extremely complicated task, we have chosen a more
conservative approach, which explores the effects of
the quantum discrete geometry underlying LQC only on the
homogeneous gravitational sector. A natural treatment
for the inhomogeneities is then a Fock quantization.
This approach assumes that there exists a regime in
which the most relevant phenomena emerging from
quantum geometry are those affecting the homogeneous
subsystem, whereas such effects are small
and can be ignored for the inhomogeneities, even if
the latter may still present a standard (Fock) quantum
behavior. Although our quantization approach is not
completely derived from LQG, it retains interesting
features associated with the discrete and polymeric
nature of the geometry, declaring a certain type of
perturbative hierarchy on their relevance for
different subsectors of the cosmological system.

As we have commented, the homogeneous sector of the
Gowdy model coincides with the phase space of the
Bianchi I model. It turns out that there exist two
different quantizations of this spacetime in the
literature of LQC, developed in
Refs. \cite{mmp} and \cite{awe1}, which correspond to
two different schemes of the so-called {\sl improved
dynamics} prescription and that, for brevity,
we will denote as
schemes A and B, respectively. In Refs.
\cite{let,ijmp}, we presented our hybrid quantization
making use of scheme A in the representation of the
homogeneous sector, i.e., applying the quantization
of Ref. \cite{mmp} (see also Ref. \cite{chio}).
The main aim of this paper is to
revisit and develop our hybrid quantization by
incorporating the alternative scheme B, put forward
by Ashtekar and Wilson-Ewing, in the
construction of the Bianchi I representation for the
homogeneous sector. We will review the quantization of
Ref. \cite{awe1}, improve and complete some aspects of
the quantization procedure, extend our hybrid framework to
adapt it to the alternative scheme and, finally,
compare the results obtained with the two hybrid
approaches.

With this objective, the constraints present in the
model will be represented by well-defined operators on
a (hybrid) kinematical Hilbert space. These
constraints are a global diffeomorphism constraint,
which generates translations in the circle, and a
global Hamiltonian constraint which couples the
homogeneous and inhomogeneous sectors in a complicated
form. As it is typical in the context of LQC, in both
schemes the Hamiltonian constraint operator {\sl
superselects} the kinematical Hilbert space ---which as a whole is
non-separable--- in separable sectors. In
particular we will be able to determine the structure
of these superselection sectors for scheme B, which
are highly non-trivial and had not been identified
previously (see Ref. \cite{awe1}). Furthermore, our
Hamiltonian constraint leads in fact to a difference
equation in an internal, strictly positive parameter
with support on semilattices of points with constant
separation. In scheme A, this parameter is the
classical global time used in the Fock quantization to
deparametrize the system \cite{men1}. In Refs.
\cite{let,ijmp} we showed that the solutions to the
Hamiltonian constraint are determined by the data on
the {\sl initial} section, namely the section where
that parameter takes its lowest allowed value. We will
see that, in scheme B, the analogous
discrete variable is (up to a constant factor) the
volume of the Bianchi I spacetime associated with the
Gowdy universe, and we will analyze the resulting
structure of the solutions.

Actually, one of the motivations of our work is the
expectation that the
quantum field theory should remain
valid on the loop quantized Bianchi I background after
imposing the quantum constraints, therefore validating
the standard Fock description for the inhomogeneities
in the approximation adopted here. In the
hybrid approach for scheme A, we demonstrated in Refs. \cite{let,ijmp}
that the physical Hilbert space has the expected tensor
product structure: the physical Hilbert space of the Bianchi I
model times a Fock space for the inhomogeneous sector,
which is equivalent to that obtained in Ref.
\cite{men1}. Hence, we indeed recover the standard
quantum field theory for the inhomogeneities. We will
discuss whether this continues to be the case in the
new hybrid approach, obtained with the alternative
quantization scheme for Bianchi I.

Another important motivation, as we have
commented, is the study of the fate of the cosmological
singularities in the quantum theory. We will see that
the polymeric quantization of the homogeneous sector,
irrespective of the adopted improved dynamics
scheme, is enough to {\sl cure} the singularity. This
fact endorses the previous results of singularity
resolution obtained in homogeneous LQC.

The rest of the paper is organized as follows. In Sec.
\ref{sec:classic} we summarize the basic features of
the classical Gowdy $T^3$ model. In Sec. \ref{sec:kin}
we construct the kinematical Hilbert space, and
represent the elementary operators in both schemes A
and B. In addition, we represent the diffeomorphism
constraint in the quantum theory. We construct the
Hamiltonian constraint operator in Sec. \ref{sec:Ham}
for the two studied schemes. In Sec. \ref{sec:physA}
we review the structure of the physical Hilbert space
for scheme A. The physical sector for case B is
investigated in Sec. \ref{sec:physB}. In Sec.
\ref{sec:results}, we discuss the main results of this
work, and conclude with some additional remarks.
Finally, we include four
appendices, which contain some supplementary technical
details.

\section{The reduced model}
\label{sec:classic}

Gowdy models describe globally hyperbolic spacetimes in vacuo
which possess two spacelike commuting Killing vector fields
and whose spatial sections are compact \cite{gowd}. The simplest
case is the Gowdy $T^3$ model (i.e., the spatial topology is
that of a three-torus). We will focus our discussion on the subsystem
with linearly polarized gravitational waves. This model
satisfies the additional restriction that the Killing
fields be hypersurface orthogonal. Hence, they can be chosen
mutually orthogonal everywhere. Let $\partial_\sigma$ and
$\partial_\delta$ be these two axial Killing fields. We may then
describe the spacetime (globally) with coordinates adapted to these
symmetries: $\{t,\theta,\sigma,\delta\}$, with
$\theta,\sigma,\delta\in S^1$. The metric components depend only
on $t$ and $\theta$. Since they are periodic in
$\theta$, they can be expanded in a Fourier series.

We use the symmetries of the system to fix completely the gauge
freedom associated with the diffeomorphism constraints in $\sigma$
and $\delta$, the directions defined by the Killing fields, as explained
in Ref. \cite{man}. We further fix all the gauge freedom associated
with the inhomogeneous (non-zero)
modes of the $\theta$-momentum constraint
and of the Hamiltonian constraint. The details of this gauge
fixing can be found in Ref. \cite{ijmp}. At the end of the day, two
global constraints remain in the model: a generator of translations
in the circle, $C_\theta$, and a Hamiltonian constraint,
$C_\text{G}$. Except for these constraints, all the gauge
freedom is fixed. The reduced phase space contains
three pairs of ``point-particle'' degrees of freedom, which
parameterize the sector of (spatially) homogenous spacetimes and
is therefore called the homogeneous sector, and all
the non-zero modes of a field and its conjugate momentum,
which form the so-called inhomogeneous sector.

In order to quantize the homogeneous sector with polymeric techniques
according to LQC, we need to describe it in terms of Ashtekar variables.
The subfamily of homogeneous solutions of the Gowdy model, encoded
in this sector as we have mentioned, corresponds to empty
Bianchi I spacetimes with three-torus topology. In a diagonal gauge, the
non-trivial components of the $SU(2)$ gravitational connection and of
the densitized triad are ${c_{i}}/{(2\pi)}$ and ${p_{i}}/{(4\pi^2)}$,
respectively, with
$\{c_i,p_j\}=8\pi G\gamma\delta_{ij}$ and $i,j=\theta, \sigma, \delta$
(see e.g. \cite{chi2,mmp,awe1}).
Here $G$ is the Newton constant and $\gamma$ the Immirzi parameter.
For the inhomogeneous sector we will carry out the Fock
quantization presented in Ref. \cite{men1}. With this aim we
adopt as basic variables the creation and annihilation variables
which would be naturally associated with a free massless scalar field,
$(a_m,a_m^*)$, where $m$ can be any non-zero integer.
In Appendix \ref{metric-new} we provide the form of
the metric in terms of the chosen variables. The constraints are
given by \cite{ijmp}
\begin{align}
C_\theta&=\sum_{m=1}^\infty
m(a_m^*a_m-a_{-m}^*a_{-m})=0,\label{ct}
\\
C_{\text{G}}&=
\frac{{\cal C}_{\text{G}}}{\sqrt{|p_\theta p_\sigma p_\delta|}}=0,
\\
{\cal C}_{\text{G}}&=-\frac{2}{\gamma^2}
\bigg[c_\theta p_\theta c_\sigma p_\sigma+
c_\theta p_\theta c_\delta p_\delta+c_\sigma p_\sigma c_\delta
p_\delta\bigg]\label{Cclas}\nonumber\\
&+G\bigg[\frac{(c_\sigma p_\sigma+
c_\delta p_\delta)^2}{\gamma^2|p_\theta|}
H_\text{int}^\xi+32\pi^2|p_\theta|
H_0^\xi\bigg],\\
\label{HIclas} H_\text{int}^\xi
&=\sum_{m\neq
0}\frac{1}{2|m|}\left[2a^*_ma_m+
a_ma_{-m}+a^*_ma^*_{-m}\right],\\\label{H0clas} H_0^\xi&=\sum_{m\neq
0}|m|a^*_ma_m.
\end{align}

The first line of ${\cal C}_{\text{G}}$ is just
the standard form of the densitized Hamiltonian constraint of the
Bianchi I model in Ashtekar variables (see e.g. \cite{chi2}). The
inhomogeneities are present in the term $H_0^\xi$, which is the
Hamiltonian corresponding to a free massless scalar field, and in
the term $H_\text{int}^\xi$, which represents an interaction term
quadratic in the field. Note that the inhomogeneities are coupled
with the homogeneous sector in a complicated way, what makes
the hybrid quantization of this system far from straightforward.

\section{The kinematical Hilbert space}
\label{sec:kin}

The kinematical Hilbert space of this reduced Gowdy model is the
tensor product of the kinematical Hilbert spaces of the two sectors.

\subsection{The homogeneous sector}

In order to construct the kinematical Hilbert space of the
homogeneous sector, we consider the two different polymeric
quantizations of the Bianchi I model developed
in Refs. \cite{mmp, awe1}. The main distinction between them
lies in the way in which the quantization prescription known
as improved dynamics (introduced in Ref. \cite{aps3} with
successful results in isotropic cosmologies) is adapted to the
anisotropic case. Although the quantization of Ref.
\cite{mmp} is well defined in the considered model with compact
spatial slices, it suffers from some drawbacks in non-compact
situations \cite{chi2}. This fact motivated the consideration
of an alternative prescription, whose corresponding quantization
has been recently developed by Ashtekar and Wilson-Ewing
in Ref. \cite{awe1}. In the following,
we will call scheme A the prescription adopted in Ref. \cite{mmp},
and scheme B that of Ref. \cite{awe1}.

Let us summarize the main characteristics of both quantizations.
The elementary configuration variables are holonomies of connections
computed along edges of oriented coordinate length $2\pi\mu_i$ in the
direction $i$, where $\mu_i$ is any real number. On the other hand,
the elementary momentum variables are triad fluxes through rectangles
orthogonal to those directions. The configuration algebra is the algebra
of almost periodic functions generated by the matrix elements of the
holonomies, namely $\mathcal N_{\mu_j}(c_j)=\exp(i\mu_{j}c_{j}/2)$
(no Einstein summation convention is adopted).
We call $\text{Cyl}_\text{S}=\otimes_i\text{Cyl}_{\text{S},i}$ the
corresponding vector space, and employ the ket notation
$|\mu_i\rangle$ to denote the states $\mathcal N_{\mu_i}(c_i)$
in the triad representation. The kinematical Hilbert space, $\mathcal
H_{\text{Kin}}=\otimes_i \mathcal H_{\text{Kin},i}$, is then the
completion of the space $\text{Cyl}_\text{S}$ with respect to
the discrete inner product
$\langle\mu_i|\mu_i^\prime\rangle=\delta_{\mu_i \mu_i^\prime}$ for
each direction. Hence, the states $|\mu_i\rangle$, which are
eigenstates of the operator $\hat p_i$ associated with fluxes,
provide an {\em orthonormal} basis of $\mathcal H_{\text{Kin},i}$.
In turn, the operators $\hat{\mathcal N}_{\mu_i'}$ associated
with holonomies produce a shift equal to
$\mu_i'$ in the label $\mu_i$
of this basis of states (see Refs. \cite{abl,chio,mmp}).
We assume that the action of any operator defined
on $\text{Cyl}_{\text{S}}$ is the identity when acting on
the kinematical Hilbert space of the inhomogeneous sector, which
will be introduced below.

\subsubsection{Improved dynamics: schemes A and B}

The so-called improved dynamics prescription is based
on the assertion that, because of the
existence of a minimum gap $\Delta$ in the spectrum of the
physical area operator in LQG \cite{aps3,note}, fiducial surfaces
cannot be as small as one wants, but there is a minimum fiducial
surface, $S_\text{min}$, whose corresponding physical area
is equal to $\Delta$. As a consequence, the coordinate length of
the holonomy along each edge turns out to exhibit a minimum
non-zero value $2\pi\bar\mu_i$. Prescription A is derived assuming
that $S_\text{min}$, lying e.g. in the $j-k$-plane, is a fiducial
square with minimum fiducial side determined by $2\pi\bar\mu_i$,
whereas according to prescription B it must be
instead a rectangle with minimum fiducial sides given by $2\pi\bar\mu_j$
and $2\pi\bar\mu_k$.
Here, and in what follows, whenever the three
indices $i$, $j$, and $k$ appear together,
we assume $\epsilon_{ijk}\neq 0$.

As a result one gets two different expressions
for $\bar\mu_i$, one for each prescription, given by
\cite{chio,chi2,awe1}
\begin{equation}\label{mubarra}
\text{A:}\quad{\frac1{\bar\mu_i}}=\frac{\sqrt{|p_i|}}{\sqrt{\Delta}},
\qquad
\text{B:}\quad{\frac1{\bar\mu_i}}=\frac1{\sqrt{\Delta}}\sqrt{\left|\frac{
p_j p_k}{p_i}\right|}.
\end{equation}

Although these arguments can be considered heuristic inasmuch as
the true relation between full LQG and LQC is still to be understood, there
is a feature of prescription B (deduced in Ref. \cite{awe1} following a
procedure of this type) which distinguishes it and has not
been realized until now in the literature. Provided that the $\bar\mu_i$'s
depend only on fluxes [i.e. $\bar\mu_i=\bar\mu_i(p_j)$],
because so does the physical area,
prescription B is uniquely determined by the requirement that,
for all directions $i$,
the exponents $\bar\mu_i c_i$ of the holonomy elements
$\mathcal N_{\bar\mu_i}(c_i)$
have a fixed {\em constant} Poisson bracket with the variable
\begin{equation}v:=\text{sgn}(p_\theta p_\sigma p_\delta)
\frac{\sqrt{|p_\theta p_\sigma p_\delta|}}{2\pi\gamma
l_{\text{Pl}}^2\sqrt{\Delta}},
\end{equation}
up to a sign depending on the orientations of the triad components.
Here, $l_{\text{Pl}}=\sqrt{G\hbar}$ is the Planck length.
Note that $v$ is proportional
to the volume of the associated Bianchi I universe. Then, at least at
the level of the Poisson bracket algebra,
the introduced requirement can be understood
as the condition that the holonomies produce a constant shift in
the volume.

For any of the prescriptions, and owing to the dependence of
$\bar\mu_i$ on the coefficients
of the densitized triad, the elementary
operator $\hat{\mathcal N}_{\bar\mu_i}$ generates in fact a
state-dependent non-linear transformation on the basis of states
$|\mu_\theta,\mu_\sigma, \mu_\delta\rangle=\otimes_i|\mu_i\rangle$.
It is possible to relabel this basis with affine parameters instead of
the labels $\mu_i$, so that the transformation generated by
$\hat{\mathcal N}_{\bar\mu_i}$ is simply a shift in the labels.
Each prescription requires a different affine reparametrization.

\subsubsection{Quantum representation in case A}

In scheme A, since $\bar\mu_i$ only depends on $p_i$,
the three fiducial directions are not mixed, and we can calculate
a new parameter $v_i(\mu_i)$ for each fiducial direction to rename
the states in our basis, so that the operator
$\hat{\mathcal N}_{\bar\mu_i}$
generates a constant shift in the new label $v_i$, as was done in
Refs. \cite{aps3,chio}. The action of the basic operators on
the relabeled states $|v_\theta,v_\sigma,v_\delta\rangle$
is given by \cite{chio,mmp}
\begin{align}
\hat{\mathcal N}_{\pm\bar\mu_i}|v_i\rangle&=|v_i\pm1\rangle,\\
\label{repA}
\hat p_i|v_i\rangle&=(6\pi\gamma l_{\text{Pl}}^2\sqrt{\Delta})^\frac{2}{3}
\text{sgn}(v_i)|v_i|^{\frac2{3}}|v_i\rangle.
\end{align}

\subsubsection{Quantum representation in case B}

Similarly, one can introduce new parameters $\lambda_i$ for the three
fiducial directions such that the action of the operator
$\hat{\mathcal N}_{\bar\mu_i}$ has only a non-trivial effect on the label
$\lambda_i$, whereas it does not change the
other two labels of the state, $\lambda_j$ and $\lambda_k$ \cite{awe1}.
Remarkably, this effect is not a constant shift anymore but depends
on the values of those two labels. Nonetheless, as we have commented,
there exists a variable $v$, given in terms of the $\lambda_i$'s by
$v=2\lambda_\theta\lambda_\sigma\lambda_\delta$, such that all
the operators $\hat{\mathcal N}_{\bar\mu_i}$ produce a constant shift
on it (for fixed orientation of the triad components). Therefore, it is
convenient to work e.g. with the
relabeled states $|v,\lambda_\sigma,\lambda_\delta\rangle$.
The two $\lambda$'s are variables which measure the degree
of anisotropy. The representation of the basic
operators is determined by \cite{awe1}
\begin{align}\label{repB}
\hat{\mathcal N}_{\pm\bar\mu_\theta}|v,\lambda_\sigma,
\lambda_\delta\rangle&=|v\pm\text{sgn}
(\lambda_\sigma\lambda_\delta),\lambda_\sigma,
\lambda_\delta\rangle,\\ \label{ptheta}
\hat p_\theta|v,\lambda_\sigma,\lambda_\delta\rangle&=
(4\pi\gamma l_{\text{Pl}}^2\sqrt{\Delta})^{\frac2{3}}
\text{sgn}
\left(\frac{v}{\lambda_\sigma\lambda_\delta}\right)
\frac{v^2}{4\lambda_\sigma^2\lambda_\delta^2}
\nonumber\\
&\times|v,\lambda_\sigma,\lambda_\delta\rangle,\\
\hat{\mathcal N}_{\pm\bar\mu_\sigma}|v,\lambda_\sigma,
\lambda_\delta\rangle&=\bigg|v\pm\text{sgn}
(\lambda_\sigma v),\lambda_\sigma\pm\left|
\frac{\lambda_\sigma}{v}\right|,
\lambda_\delta\bigg\rangle, \label{Nsigma}\\
\hat p_\sigma|v,\lambda_\sigma,\lambda_\delta\rangle&=(4\pi\gamma
l_{\text{Pl}}^2\sqrt{\Delta})^{\frac2{3}}\text{sgn}
(\lambda_\sigma)\lambda_\sigma^2|v,\lambda_\sigma,
\lambda_\delta\rangle\label{psigma}.
\end{align}
The actions of $\hat{\mathcal N}_{\pm\bar\mu_\delta}$ and
$\hat p_\delta$ can be obtained from Eqs.
\eqref{Nsigma} and \eqref{psigma} by interchanging $\lambda_\sigma$
and $\lambda_\delta$ \cite{note1}. When we construct the
Hamiltonian constraint in Sec. \ref{sec:Ham}, we will see that
states with $v=0$ ---and a fortiori with
vanishing $\lambda_\sigma$ or $\lambda_\delta$--- are removed from
our kinematical Hilbert space, and therefore the above
representation is well defined. Unlike in case A, note that
all directions are mixed now, and the operators $\hat{\mathcal
N}_{\bar\mu_i}$ and $\hat{\mathcal N}_{\bar\mu_j}$ ($i\neq j$)
do not commute.

\subsection{The inhomogeneous sector}

For this sector we employ a Fock quantization,
promoting the variables $a_m$ and $a^*_m$ to annihilation and
creation operators $\hat a_m$ and $\hat a^\dagger_m$, respectively,
such that $[\hat a_m,\hat a^\dagger_{\tilde m}]=\delta_{m\tilde m}$.
As before, we assume that these operators are the identity
acting on the homogeneous sector.
We call $\mathcal S$ the vector space
whose elements are finite linear combinations of $n$-particle
states,
\begin{equation}
|\mathfrak n \rangle:=|...,n_{-2},n_{-1},n_1,n_2,...\rangle,
\end{equation}
with $\sum_mn_m<\infty$, $n_m\in\mathbb{N}$ being the occupation number (or
number of particles) of the $m$-th mode. Then, the 
symmetric Fock space
$\mathcal F$ is the completion of the space $\mathcal S$ with
respect to the Fock inner product
$\langle\mathfrak n^\prime|\mathfrak n\rangle=\delta_{\mathfrak
n^\prime\mathfrak n}$.
Therefore the $n$-particle states provide an orthonormal basis of
the Fock space $\mathcal F$.

In the totally deparametrized system, this is the unique Fock
quantization in which the field dynamics is unitarily implemented
and that also provides a natural unitary
implementation of the gauge group of $S^1$ translations \cite{men3}.

\subsection{Quantum representation of the $S^1$ symmetry}

Since the generator of translations in the circle, given in Eq. \eqref{ct},
only affects the inhomogeneities, it is the same for the Gowdy model
in both schemes A and B. Employing the above Fock representation and
taking normal ordering we obtain its quantum
counterpart
\begin{align}
\widehat{C}_\theta=\sum_{m>0}^\infty m\hat{X}_m,\quad\hat{X}_m=
\hat{a}^{\dagger}_m
\hat{a}_m-\hat{a}^{\dagger}_{-m} \hat{a}_{-m}.
\end{align}
The $n$-particle states annihilated by this
operator are those which satisfy the condition
\begin{equation}\label{Fphys}
\sum_{m>0}^\infty m X_m=0,\quad X_m=n_m-n_{-m}.
\end{equation}
They form a dense set of a proper subspace of the Fock space
$\mathcal F$ that we denote by $\mathcal F_p$. It is (unitarily
equivalent to) the physical Hilbert space of Ref.
\cite{men1}.

\section{The Hamiltonian constraint operator}
\label{sec:Ham}

Physical states must also be annihilated by the operator that
represents the (non-densitized) Hamiltonian constraint
\cite{note2}. We will carry out a process of densitization that
will allow us to give an equivalent (and more convenient)
description in which physical states will be annihilated by the
densitized version of the Hamiltonian constraint. Actually, this
procedure is also adopted in the quantizations of Refs.
\cite{mmo,mmp}.

\subsection{Densitization of the Hamiltonian constraint}
\label{densitization}

We define the subspace of zero homogeneous volume states as the
kernel of the homogeneous volume operator
$\hat V=\otimes_i\widehat{\sqrt{|p_i|}}$ which
represents the physical volume of the (coordinate cell in the)
Bianchi I spacetime associated with the Gowdy
cosmology. Let then $\widehat C_{\text{G}}=\widehat
C_{\text{BI}}+\widehat C_{\xi}$ be the operator that represents
the (non-densitized) Hamiltonian constraint for the Gowdy model,
where $\widehat C_{\text{BI}}$ denotes the Hamiltonian constraint
for the Bianchi I model and $\widehat C_{\xi}$ is the term that
involves the inhomogeneities. One first constructs the operator
$\widehat C_{\text{BI}}$ following LQG procedures. When
symmetrizing it, one can always adopt a suitable factor ordering
such that $\widehat{C}_{\text{BI}}$ annihilates the subspace of
states with zero homogeneous volume and leaves its orthogonal
complement invariant. On the other hand, one can construct the other operator
$\widehat C_{\xi}$ so that it inherits the same properties. Therefore
the subspace of states with vanishing homogeneous
volume decouples under the action of the full Hamiltonian constraint
$\widehat C_{\text{G}}$, and one can then remove it from the
kinematical Hilbert space and restrict the study to its complement.
This complement will be denoted by
$\widetilde{\mathcal H}_\text{Kin}\otimes\mathcal F$,
where $\widetilde{\mathcal H}_\text{Kin}$ is
the completion of
\begin{align}
\widetilde{\text{Cyl}}_\text{S}&=\text{span}
\{|v_\theta,v_\sigma,v_\delta\rangle;\,
v_\theta v_\sigma v_\delta\neq0\}\nonumber\\
&= \text{span}\{|v,\lambda_\sigma,\lambda_\delta
\rangle;\,v \lambda_\sigma\lambda_\delta\neq 0\}.
\end{align}
In the last formula, we have made explicit that only
non-zero values of $\lambda_\sigma$ and $\lambda_\delta$
are allowed, even if this is implicit in the non-vanishing
of $v=2\lambda_\theta\lambda_\sigma\lambda_\delta$.

In general, non-trivial physical states, which are
annihilated by $\widehat C_{\text{G}}$, are not normalizable
in $\widetilde{\mathcal
H}_\text{Kin}\otimes\mathcal F$. In principle, they should
belong to a larger space, typically the algebraic dual
of a suitable dense subspace of this kinematical Hilbert space,
e.g. the  tensor product of $\widetilde{\text{Cyl}}_\text{S}$ and
a suitable dense subspace of $\mathcal F$. We will denote
these states by $(\tilde\psi|$. Actually, they can be
transformed into other states $(\psi|$ by the map
\begin{equation}\label{bij}
(\tilde\psi|\longrightarrow(\psi|=(\tilde\psi|
\widehat{\left[\frac{1}{V}\right]}^{\frac1{2}},
\end{equation}
which is a bijection in the considered kind of algebraic dual spaces.
Here, the operator $\widehat{[1/V]}$ represents the inverse of the
homogeneous volume and is well defined in
$\widetilde{\text{Cyl}}_\text{S}\otimes\mathcal F$. In Appendix
\ref{invV} we provide its explicit form for each of the schemes
A and B. In both cases, the resulting operator is diagonal in
our basis of states, annihilates the zero homogeneous volume
states and is bounded. Therefore it can be extended uniquely to the
kinematical Hilbert space. Moreover, its inverse
$[\widehat{1/V}]^{-1}$ is also well defined via the
spectral theorem once we have restricted the discussion to the
kinematical Hilbert space
$\widetilde{\mathcal H}_\text{Kin}\otimes\mathcal F$, because
the discrete spectrum of $\widehat{[1/V]}$ in this space does not
contain the zero anymore.

The transformed physical states $(\psi|$ are then annihilated by
the (adjoint of the) symmetric operator
\begin{align}\label{dens}
\widehat{{\cal C}}_{\text{G}}&=
\widehat{\left[\frac1{V}\right]}^{-\frac1{2}}
\widehat{C}_{\text{G}}\widehat{\left[\frac{1}{V}
\right]}^{-\frac1{2}}=\widehat{\mathcal C}_{\text{BI}}+
\widehat{\mathcal C}_\xi.
\end{align}
It is worth noting that the relation between the Hamiltonian
constraint of the Gowdy model and its densitized version does not
involve the volume of the Gowdy spacetime, but the volume of the
associated Bianchi I spacetime. Hence,
the above operator $\widehat{{\cal C}}_{\text{G}}$ is in fact the
quantum counterpart of the constraint
${\cal C}_{\text{G}}$  given in Eq.
\eqref{Cclas}.

\subsection{Quantum representation of the densitized Hamiltonian constraint}

For both schemes, we start with the (non-densitized) Hamiltonian constraint
of the Bianchi I model $\widehat{C}_{\text{BI}}$. We take
advantage of the freedom in the factor ordering to get a representation
as convenient as possible. More specifically, as mentioned above
we adopt a symmetric factor ordering which has the two following features:
i)~the same powers of
$\widehat{|p_i|}$ appear on the left and right of
every term, in a way such that $\widehat{C}_{\text{BI}}$ decouples the zero
homogeneous volume states; ii) the factors of the type
$\widehat{\sin(\bar\mu_ic_i)} \widehat{\text{sgn}(p_i)}$ [where
$\widehat{\sin(\bar\mu_jc_j)}=i(\hat{\mathcal N}_{-2\bar\mu_j}-
\hat{\mathcal N}_{2\bar\mu_j})/2$],
are symmetrized in the form
\begin{equation}\label{sym}
\frac{1}{2}\left[\widehat{\sin(\bar\mu_ic_i)}
\widehat{\text{sgn}(p_i)}+\widehat{\text{sgn}(p_i)}
\widehat{\sin(\bar\mu_ic_i)}\right].
\end{equation}
As a consequence, our operator $\widehat{C}_{\text{BI}}$ also decouples
states with different orientations of the densitized triad components.
Both properties have relevant consequences, as was discussed in
detail in Ref. \cite{mmo}, where the flat Friedmann-Robertson-Walker model
coupled to a scalar field was quantized adopting the same procedure. As we
will see later on, the procedure leads to simple superselection sectors
with neat physical properties. In particular, this fact allowed us to solve
explicitly the Hamiltonian constraint and determine the physical Hilbert
space of the hybrid Gowdy model for scheme A in Refs. \cite{let,ijmp}.
Therefore, it seems most reasonable to apply the same kind of symmetrization
in case B as well.

By densitizing $\widehat{C}_{\text{BI}}$ we obtain then the densitized
Bianchi I term $\widehat{\mathcal C}_{\text{BI}}$. The contribution of
the inhomogeneities is contained in the term $\widehat{\mathcal C}_\xi$,
which is constructed by promoting the second line of Eq. \eqref{Cclas}
to a symmetric operator. In particular, the comparison between
$\widehat{\mathcal C}_{\text{BI}}$ and its classical counterpart [first
line of Eq. \eqref{Cclas}] gives us a natural quantum prescription to
represent the term $(c_\sigma p_\sigma+c_\delta p_\delta)^2$. In addition,
we also know how to represent the term $1/|p_\theta|$, as explained in
Appendix \ref{invV}. These terms, acting on the homogeneous
sector, have a different representation in schemes A and B, and we
will analyze them for each case separately. But let us deal first with
the terms that have a non-trivial action in the inhomogeneous
sector, which are the same for both cases.

\subsubsection{Inhomogeneities}

Choosing normal ordering, the quantum counterparts of the free
Hamiltonian and the interaction term are
\begin{align}\label{H0}
\widehat{H}_0^\xi&=\sum_{m>0}^\infty m\hat N_m,\quad\hat N_m=
\hat{a}^{\dagger}_m \hat{a}_m+\hat{a}^{\dagger}_{-m} \hat{a}_{-m},
\\
\label{HI} \widehat{H}_\text{int}^\xi
 &=\sum_{m>0}^\infty\frac{\hat N_m+\hat Y_m}{m},\quad\hat Y_m=\hat{a}_m
\hat{a}_{-m}+\hat{a}^{\dagger}_m\hat{a}^{\dagger}_{-m}.
\end{align}

The inhomogeneous sector of the kinematical Hilbert space, i.e. the
Fock space $\mathcal F$, can be written as a direct sum of dynamically
invariant Fock subspaces. Indeed, the operator $\hat Y_m$, which is the
only operator in the Hamiltonian that does not act diagonally on the
basis of states $|\mathfrak n\rangle$ of $\mathcal F$, annihilates and
creates pairs of particles in modes with the same wavenumber (i.e., the modes
with wave vectors $m$ and $-m$). Therefore the quantities $X_m$, defined in
Eq. \eqref{Fphys},
are conserved under the action of the Hamiltonian constraint
$\widehat{\mathcal C}_\text{G}$. Hence, it is
convenient to relabel the basis of $n$-particle states with the quantum
numbers $X_m$, for all positive
integers $m$, together e.g. with the eigenvalues $N_m=n_m+n_{-m}$ of the
operators $\hat N_m$, defined in Eq.
\eqref{H0}. That is, we rewrite the states in our basis as
\begin{align}
|X_1, X_2,...;N_1,N_2,...\rangle:=|\mathfrak X;\mathfrak N\rangle.
\end{align}
Here, the numbers $X_m$ can take any integer value, whereas
$N_m\in\{|X_m|+2k, k\in\mathbb{N}\}$.

On these states, the action of the relevant operators is
\begin{align}\label{actX}
&\hat X_m|\mathfrak X;\mathfrak N\rangle=X_m|\mathfrak X;\mathfrak
N\rangle,\\
\label{actN} &\hat N_m|\mathfrak X;\mathfrak N\rangle=N_m|\mathfrak
X;\mathfrak N\rangle,\\\label{actF} &\hat Y_m|\mathfrak X;\mathfrak
N\rangle=\frac{\sqrt{N_m^2-X_m^2}}{2}|
\mathfrak X;...,N_m-2,...\rangle\nonumber\\
&+\frac{\sqrt{(N_m+2)^2-X_m^2}}{2}|\mathfrak
X;...,N_m+2,...\rangle,
\end{align}
and thus the sequence $\mathfrak X=\{X_1,X_2,...\}$ is not affected,
as we pointed out above.

In addition, we will denote by $\mathcal S_{\mathfrak X}$ the subspace of
$\mathcal S$ spanned by the $n$-particle states with fixed sequence
$\mathfrak X$, and by $\mathcal F_{\mathfrak X}$ the respective completion.
Then
\begin{align}
\mathcal F&=\oplus_{\mathfrak X}\mathcal F_{\mathfrak X}.
\end{align}
In practice, as far as the action of the Hamiltonian constraint operator is
concerned, we can restrict the study of the inhomogeneous sector
to any specific subspace $\mathcal F_{\mathfrak X}$. In what
follows, to simplify the notation, we will denote the $n$-particle states
by $|\mathfrak N\rangle$ and obviate their dependence in the fixed
sequence $\mathfrak X$.

Obviously the operator $\widehat{H}_0^\xi$ with domain $\mathcal S_{\mathfrak
X}$ is well defined in $\mathcal F_{\mathfrak X}$,
because it acts diagonally on the $n$-particle states and
maps $\mathcal S_{\mathfrak X}$ into itself.
On the other hand, the interaction
term $\widehat{H}_\text{int}^\xi$ creates
infinite pairs of particles, and thus it does not leave invariant the domain
$\mathcal S_{\mathfrak X}$, which only contains states with a finite number
of them. From Eqs. \eqref{HI}, \eqref{actN},
and \eqref{actF}, we have
\begin{align}\label{norm}
&||\widehat{H}_\text{int}^\xi|\mathfrak N\rangle||^2=
\bigg(\sum_{m>0}^\infty\frac{N_m}{|m|}\bigg)^2\nonumber\\
&+\sum_{m>0}^\infty\frac{N_m^2-X_m^2+2N_m}{2m^2}+\sum_{m>0}^\infty
\frac1{m^2}.
\end{align}
Since in the $n$-particle states only a finite set of occupation
numbers differ from zero, among the above three sums only the last one
involves an infinite number of non-vanishing terms. Actually this sum
converges, and hence the norm of
$\widehat{H}_\text{int}^\xi|\mathfrak N\rangle$ is finite. In
conclusion, $\widehat{H}_\text{int}^\xi|\mathfrak N\rangle \in
\mathcal F_{\mathfrak X}$ for all $|\mathfrak N\rangle\in\mathcal
S_{\mathfrak X}$, and therefore $\widehat{H}_\text{int}^\xi$ with
domain $\mathcal S_{\mathfrak X}$ is also well defined.

\subsubsection{Hamiltonian constraint in scheme A}

Following the procedure sketched in the beginning of this subsection,
for the densitized Hamiltonian constraint and
in scheme A we get an operator of the form (see Ref. \cite{ijmp}
for the details of the construction):
\begin{align}\label{CGA}
&\widehat{\mathcal{C}}_{\text{G}}^\text{A}
=\!-\frac{2}{\gamma^2}
\bigg[\widehat{\Theta}_\theta
\widehat{\Theta}_\delta+
\widehat{\Theta}_\theta\widehat{\Theta}_\sigma+
\widehat{\Theta}_\sigma
\widehat{\Theta}_\delta\bigg]\nonumber\\&+l_{\text{Pl}}^2\bigg
\{\frac{(\widehat{\Theta}_\sigma
+\widehat{\Theta}_\delta)^2}{\gamma^2}
\widehat{\bigg[\frac{1}{\sqrt{|p_\theta|}}\bigg]}^2
\!\widehat{H}_\text{int}^\xi+32\pi^2\widehat{|p_\theta|}
\widehat{H}_0^\xi\bigg\},
\end{align}
where $\widehat{H}_0^\xi$ and $\widehat{H}_\text{int}^\xi$ are given
in Eqs. \eqref{H0} and \eqref{HI}, respectively, $\widehat{|p_\theta|}$
is constructed from Eq. \eqref{repA}, and $\widehat{[1/\sqrt{|p_\theta|}]}$
is defined in Eq. \eqref{invA}. The symmetric operator
$\widehat\Theta_i$ represents the classical quantity $c_i p_i$, and its
action on the basis of states $|v_i\rangle$ of the homogeneous sector, as
determined by the quantization procedure explained above, has the
form \cite{mmp}
\begin{align}\label{acttheta}
\widehat{\Theta}_i|v_i\rangle&=-i\pi\gamma l_{\text{Pl}}^2
\big[f_+(v_i)|v_i+2\rangle-f_-(v_i)|v_i-2\rangle\big].
\end{align}
Here, $f_\pm(v_i)$ are two positive functions which satisfy that
$f_-(v_i)=f_+(v_i-2)$ and, remarkably, that $f_+(v_i)$ vanishes in the
whole interval $v_i\in[-2,0)$. Their explicit expressions are provided in
Appendix \ref{detschemeA}.

It is worth noticing that, in scheme A, the homogeneous sector is
completely factorized in three independent directional subsectors.

\subsubsection{Hamiltonian constraint in scheme B}

Motivated by our previous analysis, carried out for the Bianchi I model
\cite{mmp} as well as for the hybrid Gowdy model in scheme A
\cite{let,ijmp}, we follow in case B the very same densitization and
symmetrization procedure.

Let us focus first on the Bianchi I term. Taking into account Eq.
\eqref{mubarra} for prescription B, one can see that the densitized term
$\widehat{\mathcal C}_{\text{BI}}^\text{B}$ corresponds to the
symmetrization of
\begin{align}\label{densCBpre}
-\frac2{\gamma^2
\Delta}\sum_{j<k}\hat{V}^2\widehat{\text{sgn}(p_j)}
\widehat{\sin(\bar\mu_jc_j)}
\widehat{\text{sgn}(p_k)}\widehat{\sin(\bar\mu_kc_k)}.
\end{align}
This is precisely the gravitational part of the constraint represented in
Ref. \cite{awe1}, up to a constant multiplicative factor \cite{note3}.
Nonetheless our
symmetrization is different. Adopting a symmetrization similar to that
introduced in previous subsections, we get
\begin{align}\label{densCB}
\widehat{\mathcal C}_{\text{BI}}^\text{B}&=\widehat{\mathcal C}^{(\theta)}+
\widehat{\mathcal C}^{(\sigma)}+\widehat{\mathcal C}^{(\delta)},\\
\widehat{\mathcal C}^{(i)}&=-\frac1{4\gamma^2\Delta}\widehat{\sqrt{V}}
[\hat F_j\hat{V}\hat F_k+\hat F_k\hat{V}\hat F_j]\widehat{\sqrt{V}},\\
\hat F_i&=\widehat{\sin(\bar\mu_ic_i)}
\widehat{\text{sgn}(p_i)}+\widehat{\text{sgn}(p_i)}
\widehat{\sin(\bar\mu_ic_i)}.
\end{align}
The action of the operators $\hat{F}_i$ is displayed in Appendix
\ref{operatorF}, while the action of $\hat{V}=\otimes_i
\widehat{\sqrt{|p_i|}}$ is obtained from Eqs. \eqref{ptheta} and
\eqref{psigma}.

The final result for the densitized Hamiltonian constraint in scheme B is
\begin{align}\label{CGB}
&\widehat{\mathcal{C}}_{\text{G}}^\text{B}
=\widehat{\mathcal C}^{(\theta)}+\widehat{\mathcal C}^{(\sigma)}+
\widehat{\mathcal C}^{(\delta)}\nonumber\\&+l_{\text{Pl}}^2\bigg\{
\widehat{\bigg[\frac{1}{|p_\theta|^{\frac1{4}}}\bigg]}^2 \widehat{G}
\widehat{\bigg[\frac{1}{|p_\theta|^{\frac1{4}}}\bigg]}^2
\widehat{H}_\text{int}^\xi+32\pi^2\widehat{|p_\theta|}
\widehat{H}_0^\xi\bigg\},
\end{align}
where $\widehat{H}_0^\xi$ and $\widehat{H}_\text{int}^\xi$ are again given
in Eqs. \eqref{H0} and \eqref{HI}, $\widehat{[1/|p_\theta|^{\frac1{4}}]}$
can be found in Eq. \eqref{invB}, $\widehat{|p_\theta|}$ is constructed from
Eq. \eqref{ptheta}, and we choose the symmetric operator
\begin{align}\label{G}
\widehat{G}=\frac1{4\gamma^2\Delta}\widehat{\sqrt{V}}[\hat F_\sigma\hat{V}
\hat F_\sigma+\hat F_\delta\hat{V}\hat F_\delta]\widehat{\sqrt{V}}-
\widehat{\mathcal C}^{(\theta)}
\end{align}
to be the quantum counterpart of
$[(c_\sigma p_\sigma+c_\delta p_\delta)/\gamma]^2$.

In contrast with scheme A, the homogeneous sector is not factorized anymore
in three independent directional subsectors, since the operators $\hat F_i$
and $\hat F_j$ ($i\neq j$) do not commute.

\subsection{Superselection in the homogeneous sector}
\label{sec:super}

\subsubsection{Superselection in scheme A}
\label{sec:superA}

As discussed in Refs. \cite{let,ijmp},
$\widehat{\mathcal C}_\text{G}^\text{A}$ leaves invariant the
Hilbert subspaces $\mathcal H_{\varepsilon_i}^{\pm}$, defined as the
Cauchy completion (with respect to the discrete inner
product) of the subspaces
\begin{equation}
\text{Cyl}_{\text{S},\varepsilon_i}^{\pm}=\text{span}\{|v_i\rangle;
\,v_i\in\mathcal L_{\varepsilon_i}^\pm\},
\end{equation}
where $\mathcal L_{\varepsilon_i}^\pm$ denotes the semi-lattice of
step two defined by
\begin{equation}
\mathcal
L_{\varepsilon_i}^\pm=\{\pm(\varepsilon_i+2k);\,k\in \mathbb{N}\},
\qquad\varepsilon_i\in(0,2].
\end{equation}

Therefore, for each $\varepsilon_i$, the
kinematical Hilbert space
\begin{align}
\mathcal
H_{\vec\varepsilon}^+\otimes \mathcal F,\qquad{\rm with}\quad \mathcal
H_{\vec\varepsilon}^+=\otimes_i\mathcal H_{\varepsilon_i}^+,
\end{align}
provides a superselection sector, and the Hamiltonian constraint
operator is well defined in any of its subspaces $\mathcal
H_{\vec\varepsilon}^+\otimes\mathcal F_{\mathfrak
X}$, with dense domain
$\otimes_i\text{Cyl}_{\text{S},\varepsilon_i}^+\otimes
\mathcal S_{\mathfrak X}$. Here, we are assuming that physical
observables can distinguish between different modes and thus they do
not superselect $\mathcal F_{\mathfrak X}$.

In conclusion, in case A we can restrict the study to a kinematical
Hilbert space that is separable, and in which the quantum numbers
representing the homogeneous degrees of freedom are strictly
positive, with minimum values $\varepsilon_i$, and distributed on
cubic lattices of step 2.

\subsubsection{Superselection in scheme B}
\label{sec:superB}

Let us analyze the action of the Hamiltonian constraint operator
$\widehat{\mathcal{C}}_{\text{G}}^\text{B}$ on the
homogeneous sector component of any state
$|v^{\star},\lambda_\sigma^{\star},\lambda_\delta^{\star}\rangle\otimes
|\mathfrak N\rangle$. The explicit expression of this operator
is given in Appendix
\ref{operatorCB}. First, concerning just the homogeneous sector,
it is straightforward to see that the action of the constraint
leaves invariant
the subspace of positive densitized triad coefficients, given by
\begin{align}\label{cylpositive}
{\text{Cyl}}_\text{S}^+&=\text{span}
\{|v,\lambda_\sigma,\lambda_\delta\rangle;\,
v,\lambda_\sigma,\lambda_\delta>0\}.
\end{align} We restrict our discussion to this subspace from now on.

It is worth commenting that, whereas Ref. \cite{awe1} adopts the same
symmetrization for the powers of $\hat{V}$ that we have proposed (see
e.g. Refs. \cite{let,ijmp}), so that the states with vanishing
(homogeneous) volume are indeed decoupled, the symmetrization chosen for
the signs of the triad components differs from ours,
and therefore states with different orientations of those densitized
triad components are not decoupled. Moreover, although the discussion in Ref.
\cite{awe1} is also restricted to the space ${\text{Cyl}}_\text{S}^+$, that
restriction is incorporated there on the basis of the symmetry under parity,
but the (gravitational part of the) Hamiltonian constraint operator does not
leave invariant this domain, since it mixes states with different
orientations. In other words, it is only on the subspace of parity symmetric
(or antisymmetric) states that the restriction of the Hamiltonian constraint
to the sector of positive orientations can be defined, because the symmetry
under parity allows one to identify states with negative orientations
with a counterpart in this sector.

The Hamiltonian constraint produces shifts on the variable $v^{\star}$
which are equal to $4$ or $-4$. Nonetheless, in the considered sector,
a shift by $-4$ is possible only if $v^{\star}>4$. Then, if we define
\begin{equation}
\mathcal
L_{\varepsilon}=\{\varepsilon+4k;\, k\in \mathbb{N}\},\qquad\varepsilon\in(0,4],
\end{equation}
the action of
$\widehat{\mathcal{C}}_{\text{G}}^\text{B}$ never mixes states with values
of $v$ in different semi-lattices of this kind. Furthermore, acting on a
state $|v^{\star},\lambda_\sigma^{\star},
\lambda_\delta^{\star}\rangle\otimes|\mathfrak N\rangle$,
this operator produces new states with the following quantum numbers
$v$ and $(\lambda_a,\lambda_b)$, where the two identifications
$(a,b)=(\sigma,\delta)$ and $(a,b)=(\delta,\sigma)$ are allowed:
\begin{itemize}
\item $v=v^{\star}-4>0$
\begin{align*}
\left(\lambda_a^{\star},\frac{v^{\star}-4}{v^{\star}-2}
\lambda_b^{\star}\right), &\;
\left(\lambda_a^{\star},\frac{v^{\star}-2}{v^{\star}}
\lambda_b^{\star}\right), \\
\left(\frac{v^{\star}-2}{v^{\star}}\lambda_a^{\star},
\frac{v^{\star}-4}{v^{\star}-2}\lambda_b^{\star}
\right),&\; \left(\lambda_a^{\star},\frac{v^{\star}-4}{v^{\star}}
\lambda_b^{\star}
\right),\end{align*}
\item $v=v^{\star}$
\begin{align*}
\left(\lambda_a^{\star},\frac{v^{\star}}{v^{\star}+2}
\lambda_b^{\star}\right),
\left(\lambda_a^{\star},\frac{v^{\star}+2}{v^{\star}}
\lambda_b^{\star}\right), \\
\left(\frac{v^{\star}+2}{v^{\star}}\lambda_a^{\star},
\frac{v^{\star}}{v^{\star}+2}\lambda_b^{\star}
\right), \; \left(\lambda_a^{\star},\lambda_b^{\star}\right),
\end{align*}
and, if $v^{\star}>2$, also
\begin{align*}
\left(\lambda_a^{\star},\frac{v^{\star}-2}{v^{\star}}
\lambda_b^{\star}\right),\;
\left(\lambda_a^{\star},\frac{v^{\star}}{v^{\star}-2}
\lambda_b^{\star}\right),\\
\left(\frac{v^{\star}-2}{v^{\star}}\lambda_a^{\star},
\frac{v^{\star}}{v^{\star}-2}\lambda_b^{\star}
\right),\end{align*}
\item $v=v^{\star}+4$
\begin{align*}
\left(\lambda_a^{\star},\frac{v^{\star}+4}{v^{\star}+2}
\lambda_b^{\star}\right), \;
\left(\lambda_a^{\star},\frac{v^{\star}+2}{v^{\star}}
\lambda_b^{\star}\right), \\
\left(\frac{v^{\star}+2}{v^{\star}}\lambda_a^{\star},
\frac{v^{\star}+4}{v^{\star}+2}\lambda_b^{\star}
\right),\;\left(\lambda_a^{\star},
\frac{v^{\star}+4}{v^{\star}}\lambda_b^{\star}
\right).\end{align*}
\end{itemize}

We see that the effect caused on the $\lambda$-labels does not depend
on the reference quantum numbers $\lambda_\sigma^{\star}$ and
$\lambda_\delta^{\star}$, but only on the value of
$v^{\star}=\varepsilon+4k^{\star}$.
This dependence is through fractional factors whose denominator is two
or four units bigger or smaller than the numerator.
Therefore, it is possible to
see that, starting with $|v^{\star},\lambda_\sigma^{\star},
\lambda_\delta^{\star}\rangle\otimes|\mathfrak N\rangle$ and restricting
the consideration to the given value $v^{\star}$ of the label $v$,
the iterative action of the constraint operator leads only to states
whose quantum numbers $\lambda_a$ are of the form
$\lambda_a=\omega_\varepsilon\lambda_a^{\star}$ ($a=\sigma,\delta$),
with $\omega_\varepsilon$ belonging to the set
\begin{equation}
\mathcal
W_\varepsilon=\left\{\left(\frac{\varepsilon-2}{\varepsilon}
\right)^z·\prod_{m,
n\in\mathbb{N}}
\left(\frac {\varepsilon+2m } {\varepsilon+2n}\right)^{k_n^m} \right\} ,
\end{equation}
where $k_n^m\in\mathbb{N}$, and
$z\in\mathbb{Z}$ if $\varepsilon>2$, while $z=0$ when $\varepsilon\leq2$.
The discrete set
$\mathcal W_\varepsilon$ is countably infinite and turns out to be dense
in the positive real line. The proof of this last statement can be found
in Appendix \ref{proof}. Therefore, whereas the variable $v$ has
support on simple semilattices of constant step,
the variables $\lambda_a$ take values in much more complicated sets.
Nonetheless, they are also superselected in separable sectors, a fact
which had not been realized in previous literature. As a particular
case, we can see that if $\varepsilon=2$ and $\lambda_a^{\star}$ is a
fraction, then $\lambda_a$ can take any value in the set of positive
rational numbers.

In conclusion,
$\widehat{\mathcal{C}}_{\text{G}}^\text{B}$ leaves
invariant the Hilbert subspaces
$\mathcal H_{\varepsilon,\lambda_\sigma^{\star},\lambda_\delta^{\star}}$
defined as the Cauchy completion with respect to the discrete inner
product of
\begin{align}
\text{Cyl}_{\text{S},\varepsilon,\lambda_\sigma^{\star},
\lambda_\delta^{\star}}=\text{span}\{&|v,
\lambda_\sigma,\lambda_\delta\rangle;
v\in\mathcal L_{\varepsilon},\lambda_a=\omega_\varepsilon\lambda_a^{\star},
\nonumber\\&
\omega_\varepsilon\in\mathcal W_\varepsilon,\lambda_a^{\star}\in\mathbb{R}^+\}.
\end{align}
As a consecuence, the Hilbert subspaces
\begin{align}
\mathcal H_{\varepsilon,\lambda_\sigma^{\star},\lambda_\delta^{\star}}
\otimes\mathcal F
\end{align}
provide superselection sectors. Moreover, the Hamiltonian constraint operator
$\widehat{\mathcal{C}}_{\text{G}}^\text{B}$ has well-defined restrictions
on any of the subspaces $\mathcal H_{\varepsilon,\lambda_\sigma^{\star},
\lambda_\delta^{\star}}\otimes\mathcal F_\mathfrak{X}$, with corresponding
dense domain given by
$\text{Cyl}_{\text{S},\varepsilon,\lambda_\sigma^{\star},
\lambda_\delta^{\star}}\otimes\mathcal{S}_\mathfrak{X}$.

\section{Physical sector in case A}
\label{sec:physA}

\subsection{Imposition of the Hamiltonian constraint}

In Ref.\cite{ijmp} we already determined the solutions $\big(\psi\big|$
of the Hamiltonian constraint $\widehat{\mathcal C}_\text{G}^\text{A}$,
which were given by an expansion of the form
\begin{align}
&(\psi|=\!\sum_{v_\theta\in\mathcal
L_{\varepsilon_\theta}^+}\!\int_{\mathbb{R}^2}d\omega_\sigma
d\omega_\delta\langle
v_\theta|\otimes\langle\omega_\sigma|\otimes\langle\omega_\delta|\otimes(
\psi_{\omega_\sigma, \omega_\delta}(v_\theta)|.
\end{align}
Here, $|\omega_a\rangle$ ($a=\sigma,\delta$) denotes the generalized
eigenstates (with real generalized eigenvalue $\omega_a$) of
$\widehat{\Theta}_a$. These operators are constants of motion
(they commute with $\widehat{\mathcal C}_\text{G}^\text{A}$), and therefore
the states $|\omega_a\rangle$ are stable under the action of
$\widehat{\mathcal C}_\text{G}^\text{A}$. The concrete expression of
these states can be found in Ref. \cite{mmp}. On the other hand, remarkably,
the projections $(\psi_{\omega_\sigma, \omega_\delta}(\varepsilon_\theta+2k)|$
of the solution on the sections with constant value of $v_{\theta}=\varepsilon_\theta+2k$ for
every
$k\in\mathbb{N}^+$, are all formally determined by the initial data
$(\psi_{\omega_\sigma, \omega_\delta}(\varepsilon_\theta)|$ through the
action of a complicated operator, which involves the iterative action of
$\widehat{H}_0^\xi$ and $\widehat{H}_\text{int}^\xi$. Again, the explicit
expression is provided in Ref. \cite{ijmp}.

In addition, the physical solutions must be annihilated by the generator
of translations in $S^1$. This $S^1$ symmetry is preserved by the dynamics,
since the quantities $X_m$ are constants of motion; therefore we can ignore
it at this stage and impose it after dealing with the Hamiltonian constraint.

\subsection{Physical Hilbert space}

The resulting form of the general solutions $(\psi|$
of the Hamiltonian constraint is only formal, in the sense that
they do not belong to the dual
of the domain of definition of $\widehat{\mathcal C}_\text{G}^\text{A}$.
Namely, some of the coefficients
$(\psi_{\omega_\sigma,\omega_\delta}(v_\theta)|\mathfrak N\rangle$ of
the solutions diverge, for example when
$v_\theta\geq\varepsilon_\theta+4$.
Indeed, it is not difficult to see that the space $\mathcal S_\mathfrak X$
is not in the domain of the operator
$\widehat{H}_0^\xi\widehat{H}_\text{int}^\xi$, because this contains the
term $\sum_m\hat{N}_m\hat{Y}_m$. The action of this term on a
generic state of $\mathcal S_\mathfrak X$ does not lead to a
normalizable state of the Fock space $\mathcal F_\mathfrak X$, not
even in the (conventional) generalized sense, since, in particular, it
involves the creation of pairs in an infinite number of modes
and then a sum over the number of all of those particles.

This problem can be traced back to the choice of domain for the Hamiltonian
constraint, since the present one is not invariant under the action of
$\widehat{\mathcal C}_\text{G}^\text{A}$. Indeed, it is worth pointing out that the
determination of an invariant domain
(and such that the Hamiltonian constraint be an essentially self-adjoint
operator) would allow one to resort to group averaging techniques
\cite{group} in order to construct the space of physical solutions.
However, the selection of an alternative domain that remains
invariant is an extremely difficult task, given the complexity of our model,
and no satisfactory choice is actually at hand.
In this sense, the kinematical structure of our quantization is not
well adapted to the physical one, and their relation is not
straightforward a priori.

Nonetheless, we can still complete the quantization
program: we just need to make sense of the solutions and provide
them with a Hilbert structure. As we have already said, the solutions
to the densitized Hamiltonian constraint are completely determined,
at least formally, by a single piece of
initial data $(\psi_{\omega_\sigma,\omega_\delta}
(\varepsilon_\theta)|$, and then we can identify these solutions with the
corresponding data. The determination of a complete set of real classical
observables acting on these initial data,
together with the condition that they be represented as
self-adjoint operators, determines a unique inner product
\cite{red} that characterizes the Hilbert structure.
Moreover, taking into account the additional $S^1$ symmetry,
implemented by condition \eqref{Fphys}, we
conclude \cite{let,ijmp} that the initial data
$(\psi_{\omega_\sigma,\omega_\delta}
(\varepsilon_\theta)|$ must belong to the Hilbert space
\begin{align}\label{Hinitial}
L^2(\mathbb{R}^2,d\omega_\sigma d\omega_\delta)\otimes\mathcal F_p,
\end{align}
that we identify as the physical Hilbert space.

Alternatively, we can argue that this is the physical Hilbert space
following a different line of reasoning, based on the idea that we
can regularize the theory by means of a cutoff for the wavenumber $m>0$
and analyze the limit of arbitrarily large values of the cutoff.
Note that we are allowed to carry out this further reduction of the phase
space because, in the Gowdy model, the field modes with different
wavenumbers are not mixed dynamically.

\subsubsection{Regularized model}

Let $\mathcal S_{M}$ be the subspace of $\mathcal S$ spanned by the
$n$-particle states which satisfy the condition $N_m=0$ for all
$m>M>0$, and let $\mathcal S_{M}^\bot$ be its orthogonal
complement. In addition, let $\widehat P_{M}$ and
$\widehat P_{M}^\bot$ be the projectors on the
subspaces $\mathcal S_{M}$ and $\mathcal S_{M}^\bot$, respectively.
We are now interested in finding
the physical Hilbert space of the truncation of the Gowdy model in
which all the field modes with wavenumber bigger than the cutoff $M$
vanish. Obviously, this truncated system is
governed by the constraints
\begin{align}\label{proj}
&\big(\psi_{\omega_\sigma, \omega_\delta}(v_\theta)|\widehat
P_{M}^\bot=0,\\\label{projCG} &\big(\psi_{\omega_\sigma,
\omega_\delta}(v_\theta)|\widehat P_{M}\,\widehat{\mathcal
C}_\text{G}^\text{A}\,\widehat P_{M}=0,\\\label{projCtheta}
&\big(\psi_{\omega_\sigma, \omega_\delta}(v_\theta)|\widehat
P_{M}\,\widehat{\mathcal C}_\theta\,\widehat P_{M}=0.
\end{align}

Simultaneous solutions to the first two equations have
the same form as the solutions to the Hamiltonian constraint
in the full Gowdy model, but containing exclusively wavenumbers
$m$ equal or smaller than $M$, so that they now possess only
a finite number of
terms. As a consequence, the divergences caused by the infinite
production of pairs of particles disappear, and the coefficients
$(\psi_{\omega_\sigma,\omega_\delta}(v_\theta)|\mathfrak N\rangle$
of the solutions to Eqs. \eqref{proj} and \eqref{projCG}
are now finite for all $v_\theta$. Therefore, the solutions (for fixed
$v_\theta$, $\omega_\sigma$, and $\omega_\delta$) live in the
dual $\mathcal S_M^*$ of the space spanned by the $n$-particle
states with the cutoff. Similarly to what we did in the previous
subsection, we can now endow the data at $v_\theta=\varepsilon_\theta$
with a Hilbert structure. We construct the same complete set of
observables \cite{ijmp}, with the difference that now we do not have
an infinite number of them in the inhomogeneous sector, but only
$4M$ (corresponding to two degrees of freedom on phase space for
each of the $2M$ wave vectors). The resulting observables act on the
solutions $(\psi_{\omega_\sigma,\omega_\delta}(\varepsilon_\theta)|$
and are self-adjoint in $L^2(\mathbb{R}^2,d\omega_\sigma
d\omega_\delta) \otimes\mathcal F_M$, where $\mathcal F_M$ is the
Fock space obtained by completing $\mathcal S_M$. Taking into
account the remaining constraint \eqref{projCtheta}, which implements
the $S^1$ symmetry, the physically admissible states have finally
the Hilbert structure
\begin{align}\label{HphysM}
L^2(\mathbb{R}^2,d\omega_\sigma d\omega_\delta) \otimes(\mathcal
F_p){}_M,
\end{align}
where $(\mathcal F_p){}_M$ is the subspace of $\mathcal F_M$ spanned
by the $n$-particle states that verify the condition $\sum_{m>0}^M m
X_m=0$.

As the cutoff $M$ increases, we get closer to the
non-truncated theory for the Gowdy model. This indicates that,
in the limit $M\rightarrow\infty$, we should recover the full Fock
space: $(\mathcal F_p){}_M\rightarrow\mathcal F_p$. This supports
our previous statement that the physical Hilbert space of the
non-truncated Gowdy model is in fact the space
\eqref{Hinitial}.

The introduction of the cutoff in the wavenumber also makes
manageable the numerical study of the effect that the inhomogeneities
have on the dynamical behavior of the system and, in particular,
of the changes that occur in the dynamics of the Bianchi I background
when the inhomogeneities grow. Actually, this kind of analysis has
already been carried out at the effective level and is included in the
study of Ref. \cite{eff}, where the classical Hamiltonian has been modified
in an effective way in order to take into account the corrections that arise
from the quantum theory ---although the results obtained there
are extended to the full model without the cutoff.

\section{Physical sector in case B}
\label{sec:physB}

\subsection{Imposition of the Hamiltonian constraint}

Unlike in case A, the homogeneous sector of the kinematical Hilbert space,
$\mathcal H_{\varepsilon,\lambda_\sigma^{\star},\lambda_\delta^{\star}}$,
is not factorized in three independent directional subsectors, and none
of the operators
$\widehat{\sqrt{V}}\hat{F}_i\widehat{\sqrt{V}}$ appearing in the
expression of the constraint $\widehat{\mathcal{C}}_{\text{G}}^\text{B}$
(i.e., the counterpart of $\hat{\Theta}_i$ in case A) is now a constant of
motion \cite{note4}. Therefore, in this case we cannot simplify
the action of the Hamiltonian constraint operator by diagonalizing it in
the directional subsectors labeled by $\sigma$ and $\delta$, as we did
in scheme A.

Then, in order to solve the Hamiltonian constraint represented by
$\widehat{\mathcal C}_\text{G}^\text{B}$,
we expand the solutions $(\psi|$ using the basis
of states $|v,\lambda_\sigma,\lambda_\delta\rangle$ for the homogeneous
sector. Namely,
\begin{align}
(\psi|&=\sum_{v\in\mathcal
L_{\varepsilon}}\sum_{\omega_\varepsilon\in\mathcal
W_\varepsilon}\sum_{\bar\omega_\varepsilon\in
\mathcal W_\varepsilon}\langle
v,\omega_\varepsilon\lambda_\sigma^\star,\bar\omega_\varepsilon
\lambda_\delta^\star|\nonumber\\&\otimes(
\psi(v,\omega_\varepsilon\lambda_\sigma^\star,
\bar\omega_\varepsilon\lambda_\delta^\star)|.
\end{align}

Inserting this expansion into the (dual of the)
constraint equation, projecting on the
homogeneous sector, and taking into account the action of
$\widehat{\mathcal{C}}_{\text{G}}^\text{B}$ given in Appendix
\ref{operatorCB}, we get that the projections
\[(\psi(v,\lambda_\sigma,\lambda_\delta)|=
(\psi(v,\omega_{\varepsilon}\lambda_\sigma^{\star},
\bar\omega_{\varepsilon}\lambda_\delta^{\star})|\] satisfy a
series of relations that can be interpreted as
difference equations in $v$. Introducing the
projections of $(\psi|$ on the combinations of states defined
in Eqs. \eqref{state1}-\eqref{state4} of
Appendix \ref{operatorCB}, which we call
\[(\psi_\pm(v\pm4,\lambda_\sigma,\lambda_\delta)|
=(\psi|v\pm4,\lambda_\sigma,\lambda_\delta\rangle_\pm\]
and in a similar way for the rest of projections, the relation
obtained is
\begin{align}\label{solutionB}
(\psi_+&(v+4,\lambda_\sigma,\lambda_\delta)|-\beta
b_\theta^2(v,\lambda_\sigma,\lambda_\delta)b
_\theta^2(v+4,\lambda_\sigma,\lambda_\delta)\nonumber\\
\times&\frac{
v+4}{v}(\psi{}_+'(v+4,\lambda_\sigma, \lambda_\delta)
|\widehat{H}_\text{int}^\xi\nonumber\\
&=-\frac{1}{\beta}\frac{32v^2}{\lambda_\sigma^2
\lambda_\delta^2x_+(v)}(\psi(v,
\lambda_\sigma,\lambda_\delta)|\widehat{H}_0^\xi
\nonumber\\&+\frac{x^-_0(v)}{x_+(v)}(\psi_{0^-}
(v,\lambda_\sigma,\lambda_\delta)|+
\frac{x^+_0(v)}{x_+(v)}(\psi_{0^+}
(v,\lambda_\sigma,\lambda_\delta)|
\nonumber\\&-\frac{x_-(v)}{x_+(v)}
(\psi_-(v-4,\lambda_\sigma,\lambda_\delta)|
+\beta b_\theta^2(v,\lambda_\sigma,
\lambda_\delta)\nonumber\\&\times\bigg\{
b_\theta^2(v-4,\lambda_\sigma,\lambda_\delta)
\frac{v-4}{v}\frac{x_-(v)}{x_+(v)}(\psi_-'(v-4,\lambda_\sigma,
\lambda_\delta)|\nonumber\\
&-b_\theta^2(v,\lambda_\sigma,\lambda_\delta)
\bigg[\frac{x_{0}^{-}(v)}{x_+(v)}(\psi'_{0^-
}(v,\lambda_\sigma,\lambda_\delta)|\nonumber\\&+
\frac{x_{0}^{+}(v)}{x_+(v)}(\psi'_{0^+
}(v,\lambda_\sigma,\lambda_\delta)|\bigg]\bigg\}
\widehat{H}_\text{int}^\xi.
\end{align}
Here, $b_{\theta}$, $x_{\pm}$, and $x_{0}^{\pm}$ are the functions
defined in Appendices \ref{invVB} and \ref{operatorCB}.

\subsection{Analysis of the solutions}

We want to prove now that, as in case A, the solution is
completely determined by the data on the first
section $v=\varepsilon$ (at least formally).
Specifically, we want to show that, given a
set of initial data
$(\psi(\varepsilon,\omega_\varepsilon\lambda_\sigma^\star,
\bar\omega_\varepsilon\lambda_\delta^\star)|$
(belonging to $\mathcal{S}^*_\mathfrak{X}$ for all
$\omega_{\varepsilon},\bar\omega_\varepsilon\in\mathcal W_{\varepsilon}$),
it is possible to determine each term
$(\psi(v,\lambda_\sigma,\lambda_\delta)|$ of the solution, for every
$v>\varepsilon$ in ${\cal L}_{\varepsilon}$.

The presence of the interaction term on the left hand side
of Eq. \eqref{solutionB} complicates a direct proof of the above statement.
However, it is possible to attain the result by means of an asymptotic
analysis of the solutions. Remarkably, our theory
involves a dimensionless parameter, $\beta$, introduced in Eq. \eqref{beta},
and recurring to it we can naturally adopt an asymptotic approach without
the need to introduce any external parameter by hand. Note that, since the
area gap $\Delta$ is proportional to $\gamma l_{\text{Pl}}^2$, $\beta$ is
proportional to the inverse of the Immirzi parameter $\gamma$. Thus,
in the limit $\beta\rightarrow 0$, we expand the solutions in asymptotic
series of the form:
\begin{align}\label{exp}
&(\psi(\varepsilon+4k,\lambda_\sigma,\lambda_\delta)|
=\sum_{n\in\mathbb{N}}\beta^{n-k}\left({}
^n\psi(\varepsilon+4k, \lambda_\sigma ,
\lambda_\delta)\right|,\nonumber\\& \forall k\in\mathbb{N}^+.
\end{align}
Note that the linear combinations introduced in Eq. \eqref{solutionB},
like e.g. $(\psi_\pm(v\pm4,\lambda_\sigma,\lambda_\delta)|$,
adopt then similar expansions,
and we will denote their terms using an obvious notation,
for instance $({}^n\psi_\pm(v\pm4,\lambda_\sigma,\lambda_\delta)|$.
Substituting the expansion \eqref{exp} in the constraint \eqref{solutionB},
and considering powers of $\beta$ order by order, we obtain an expression
for every term $\left({}^n\psi_+(v+4,\lambda_\sigma,\lambda_\delta)\right|$
(for generic $v$) provided that the data at $v$ and $v-4$ are known.
The explicit result is the following (to simplify the notation, we obviate
the dependence of the states and of the function $b_\theta$ on the
$\lambda$'s):
\begin{itemize}
 \item Leading term:
\begin{align}\label{0}
 \big({}^0\psi_+(v+4)\big|&=-\frac{32v^2}{
\lambda_\sigma^2\lambda_\delta^2
x_+(v)}\big({}^0\psi(v)\big|\widehat{H}_0^\xi,
\end{align}
\item first order correction:
\begin{align}\label{1}
\big({}^1\psi_+&(v+4)\big|=-\frac{32v^2}{\lambda_\sigma^2
\lambda_\delta^2
x_+(v)}\big({}^1\psi(v)\big|\widehat{H}_0^\xi\nonumber\\&
+\frac {x^-_{0}(v)}{x_+(v)}
\big({}^0\psi_{0^-}(v)\big|+\frac{x^+_{0}(v)}{x_+(v)}
\big({}^0\psi_{0^+}(v)\big|\nonumber\\&
+b_\theta^2(v)b_\theta^2(v+4)\frac{
v+4}{v}\big({}^0\psi'_+(v+4)\big|\widehat{H}_\text{int}^\xi,
\end{align}
\item second order correction:
\begin{align}\label{2}
\big({}^2\psi_+&(v+4)\big|=-\frac{32v^2}{\lambda_\sigma^2
\lambda_\delta^2
x_+(v)}\big({}^2\psi(v)\big|\widehat{H}_0^\xi\nonumber\\&
+\frac{x^-_{0}(v)}{x_+(v)}\big({}^1\psi_{0^-}(v)\big|+
\frac{x^+_{0}(v)}{x_+(v)}\big({}^1\psi_{0^+}
(v)\big|\nonumber\\&-\frac{x_-(v)}{x_+(v)}
\big({}^0\psi_-(v-4)\big|+b_\theta^2(v)
\bigg\{-b_\theta^2(v)\nonumber\\&\times\bigg[\frac{
x^-_0(v)}{x_+(v)}\big({}^0\psi'_{0^-}(v)\big|+\frac{
x^+_0(v)}{x_+(v)}\big({}^0\psi'_{0^+}(v)\big|\bigg]\nonumber\\&
+b_\theta^2(v+4)\frac{v+4}{v}\big({}^1\psi'_+(v+4)
\big|\bigg]\bigg\}\widehat{H}_\text{int}^\xi,
\end{align}
\item $n$-th order correction ($n\geq3$):
\begin{align}\label{n}
\big({}^n\psi_+&(v+4)\big|=-\frac{32v^2}{\lambda_\sigma^2
\lambda_\delta^2
x_+(v)}\big({}^n\psi(v)\big|\widehat{H}_0^\xi\nonumber\\&
+\frac{x^-_{0}(v)}{x_+(v)}\big({}^{n-1}\psi_{0^-}(v)\big|+
\frac{x^+_{0}(v)}{x_+(v)}\big({}^{n-1}
\psi_ {
0^+}(v)\big|\nonumber\\&-\frac{x_-(v)}{x_+(v)}\big({}^{n-2}
\psi_-(v-4)\big|+b_\theta^2(v)\bigg\{
-b_\theta^2(v)\nonumber\\&\times
\bigg[\frac{
x^-_0(v)}{x_+(v)}\big({}^{n-2}\psi'_{0^-}(v)\big|+\frac{
x^+_0(v)}{x_+(v)}\big({}^{n-2}\psi'_{0^+}(v)\big|\bigg]
\nonumber\\&+
b_\theta^2(v+4)\frac{v+4}{v}\big({}^{n-1}\psi'_+(v+4)\big|
\nonumber\\&
+b_\theta^2(v-4)\frac{v-4}{v}\big({}^{n-3}\psi'_-(v-4)
\big|\bigg\}\widehat{H}_\text{int}^\xi.
\end{align}
\end{itemize}

The above expressions simplify considerably in the cases $v=\varepsilon$
and $v=\varepsilon+4$. This is due to the fact that, on the one hand,
the data on the very initial
section are not given by asymptotic series, that is, in principle
$(\psi(\varepsilon)|=({}^0\psi(\varepsilon)|$ and hence
$({}^n\psi(\varepsilon)|=0$ for all $n\geq1$. And, on the other hand,
$x_-(\varepsilon)=0$ for all $\varepsilon\in(0,4]$, whereas
$x^-_{0}(\varepsilon)=0$ if $\varepsilon\leq2$ [see Eqs.
\eqref{coefficient1} and \eqref{coefficient3}].

In view of these equations, we see that, for $v>4$, the knowledge of the
solution on the sections $v-4$ and $v$, together with the terms of asymptotic
order $n-1$ on the section $v+4$, determine on this last section the terms
of the solution at order $n$ via the following linear combinations
\begin{align}
({}^n\psi_+(v+4,\lambda_\sigma,\lambda_\delta)|
&=\bigg({}^n\psi\bigg(v+4,\lambda_\sigma,\frac{v+4}
{v+2}\lambda_\delta\bigg)\bigg|\nonumber\\&+\bigg({}^n\psi
\bigg(v+4,\lambda_\sigma, \frac{v+2 }
{v}\lambda_\delta\bigg)\bigg|\nonumber\\&+\bigg({}^n\psi
\bigg(v+4,\frac{v+4 }
{v+2}\lambda_\sigma,\frac{v+2}
{v}\lambda_\delta\bigg)\bigg|\nonumber\\&+\bigg({}^n\psi
\bigg(v+4,\frac{v+4}
{v+2}\lambda_\sigma,\lambda_\delta\bigg)\bigg|
\nonumber\\&+\bigg({}^n\psi\bigg(v+4, \frac{v+2}{v}
\lambda_\sigma,\frac{v+4}{v+2}\lambda_\delta\bigg)\bigg|
\nonumber\\&+\bigg({}^n\psi\bigg(v+4,\frac{
v+2}{v}\lambda_\sigma,\lambda_\delta\bigg)\bigg|.
\end{align}
On the other hand, for $0<v<4$, the data on the section $v-4$
are spurious, and their knowledge its not required to determine the
terms of the solution for $v+4$.

Actually, a similar structure appears also in the solutions to the
Hamiltonian constraint for the Bianchi I model, inasmuch as the solution
on the section $v+4$ is determined in terms of the same kind of linear
combinations. Furthermore, it has been recently shown \cite{mmw} that,
for all $v>0$, the set of linear combinations
\begin{align*}
\{\left(\psi_+(v+4,\omega_{\varepsilon}\lambda_\sigma^{\star},
\bar\omega_\varepsilon\lambda_\delta^{\star})\right|;\;
\omega_{\varepsilon},\bar\omega_\varepsilon\in\mathcal W_{\varepsilon},\}
\end{align*}
determines the set of individual terms
\begin{align*}
\{\left(\psi(v+4,\omega_{\varepsilon}\lambda_\sigma^{\star},
\bar\omega_\varepsilon\lambda_\delta^{\star})\right|;\;
\omega_{\varepsilon},\bar\omega_\varepsilon\in\mathcal W_{\varepsilon}\}
\end{align*}
through a one-to-one map. This result can be applied
here as well as for Bianchi I cosmology.
Therefore, starting with the initial data, we can obtain, step by step,
the terms of the solution ---up to the desired asymptotic order--- on all
of the consecutive $v$-sections.

In conclusion, the initial data
$(\psi(\varepsilon,\lambda_\sigma,\lambda_\delta)|$ (with
$\lambda_\sigma$ and $\lambda_\delta$ taking all possible values in
the corresponding superselection sectors) completely determines the
solution, as we wanted to show. The solutions, constructed in this way,
are formal, as happens to be the case in scheme A, in the sense that the
objects $({}^n\psi(v+4,\lambda_\sigma,\lambda_\delta)|$ do not belong
in general to the dual space of the domain chosen to define the
Hamiltonian constraint operator, owing to the presence of the operator
$\widehat{H}_\text{int}^\xi$ in their expressions.

\subsection{Physical Hilbert space}

In order to provide the solutions with a Hilbert structure, we proceed in
the very same way as in case A. Once we have shown that the set of
initial data
\begin{align*}
\{\left(\psi(\varepsilon,\omega_{\varepsilon}\lambda_\sigma^{\star},
\bar\omega_\varepsilon\lambda_\delta^{\star})\right|;\;
\omega_{\varepsilon},\bar\omega_\varepsilon\in\mathcal W_{\varepsilon}\}
\end{align*}
characterizes the solution, we can identify solutions with their
corresponding data, and the physical Hilbert space with
a Hilbert space of such initial data.

The requirement that a complete set of observables acting on these
initial data be self-adjoint operators determines again uniquely
the inner product that provides the Hilbert structure.
Such observables are given, for instance, by
a complete set of observables for the Bianchi I model in vacuo
and by the observables introduced in scheme A for the inhomogeneities.
Imposing the remaining $S^1$ symmetry on the resulting Hilbert space,
we finally get the same structure found in case A for the physical Hilbert
space, namely, the tensor product of the Fock
subspace $\mathcal F_p$ and the physical Hilbert space of the
Bianchi I model in vacuo, though now in scheme B:
\begin{align}\label{HphysB}
\mathcal H^\text{B}_{\text{phys}}=\mathcal H^\text{B}_{\text{phys,BI}}
\otimes\mathcal F_p.
\end{align}
The explicit form of $\mathcal H^\text{B}_{\text{phys,BI}}$ is
analyzed in Ref. \cite{mmw}.

\section{Summary and conclusion}
\label{sec:results}

\subsection{Recovery of the standard quantum field theory}
\label{sec:QFT}

As we have commented in the Introduction, one of the
motivations of this work is to investigate the plausibility
of the recovery of standard quantum field theory in the
framework of loop quantization. In particular we wanted to
show that, in the Gowdy model, one attains a Fock
description of the inhomogeneities over a polymerically quantized
Bianchi I background in the space of physical states, starting with a
hybrid quantization in the kinematical setting. Indeed, we have proved
that this is the case, since the physical Hilbert space obtained
in both schemes has the structure of the tensor product of the physical
Hilbert space of the Bianchi I model and a Fock space, which turns out
to be equivalent to the space obtained in the standard Fock
quantization \cite{men1,men3}. This result supports the validity of the
hybrid quantization, because the latter should lead to the standard
quantization
of the system in the limit in which the effects arising from the
discreteness of the geometry become negligible.
Let us remark that the result is non-trivial, inasmuch as the hybrid
approach is introduced in the kinematical arena and the relation between
the kinematical and physical Hilbert structures cannot be anticipated
before completing the quantization, even more if one takes into account
the field-like complexity of the model.

\subsection{Resolution of the cosmological singularity}
\label{sec:sing}

The classical solutions of the linearly polarized Gowdy $T^3$ model
present generically a cosmological singularity. In Ref.
\cite{mon}, e.g., a curvature invariant was explicitly calculated and
proven to diverge almost everywhere at initial time. In terms of the
variables that we have employed, the cosmological singularity
corresponds to vanishing values for the components $p_i$ of
the densitized triad. Actually, as we can see in Eq.
\eqref{newmetric}, the metric is ill defined if any of the $p_i$'s is
zero.

In our quantum theory, the polymeric quantization performed in the
homogeneous sector succeeds in eliminating the singularity. More
explicitly, we have been able to remove the kernel
of all the operators $\hat p_i$ and, as a consequence, an analog
of the classical cosmological singularity does not exist any more
quantum mechanically. This resolution of the singularity is
achieved at a kinematical level.
Of course, it persists in the physical Hilbert space,
since physical states do not have projection onto the zero
eigenspaces of the operators $\hat p_i$.
Furthermore, they only have support on a
sector with fixed orientation of the triad components and, then, they do not cross
the singularity to another branch of the universe
corresponding to a different orientation.

On the other hand, in addition to this kinematical resolution of the
cosmological singularity, it is worth commenting that, at least for scheme A
and in the framework of the effective description corresponding to the
hybrid quantization put forward here, the numerical simulations
performed so far for the Gowdy model show the presence of a bounce
which replaces the singularity and which emerges owing to the quantum geometry
corrections to Einstein's theory \cite{eff}. Similar numerical calculations
are being developed currently for scheme B \cite{mpw}, in order to
validate this stronger result about the resolution of the singularity.

Let us emphasize that the standard (non-polymeric) quantum methods
do not succeed in resolving the cosmological
singularity. On the one hand, in the
Fock quantization of the deparametrized system, which has
been discussed in the literature \cite{qGow} and which has
been successfully accomplished till completion in Refs. \cite{men1,men3},
a classical time parameter is present explicitly in the
quantum description, and the curvature invariant
calculated in Ref. \cite{mon} depends on its inverse in
such a way that the invariant still blows up at initial time.
On the other hand, if one does not deparametrize the system,
following our gauge reduction, and quantizes the homogeneous sector
in a standard (non-polymeric) way, as in the Wheeler-De Witt approach,
then the zero eigenvalue would be included in the continuous spectrum
of the triad operator, instead of the discrete spectrum, and there would
not exist a proper subspace associated with this eigenvalue which
could be decoupled and removed.

\subsection{Concluding remarks}
\label{sec:con}

In conclusion, we have rigorously constructed a hybrid quantization
of the Gowdy model with three-torus topology and linearly polarized
gravitational waves. The homogeneous sector of the phase space,
which coincides with the phase space of a Bianchi I model, has
been polymerically quantized, whereas we have applied a
(distinguished) Fock quantization to represent the inhomogeneous
sector. In the LQC literature, there exist two different schemes for
the polymeric quantization of the Bianchi I universes, denoted in this
paper as schemes A and B. We had already analyzed the hybrid
quantization of the Gowdy model adopting scheme A in Refs.
\cite{let,ijmp}. Here, we have revisited that quantization
and extended our hybrid approach to the alternative case B,
in which the homogeneous sector has a different representation.

In both schemes, the quantum Hamiltonian constraint has been
densitized, in order to deal with a simpler constraint, and then
has been promoted to an operator, well defined in some dense domain
of the kinematical Hilbert space. This is truly a non-trivial result,
because our system possesses an infinite number of degrees of freedom
and the two sectors, on which the constraint operator acts,
are coupled and quantized with entirely different methods.

As we have seen,
the kinematical structure over which we have
defined the theory and the choice of domain for our quantum operators do
not suffice to make sense of the formal solutions to the Hamiltonian
constraint. Nonetheless, we have found a procedure to overcome
the problem and complete the quantization.
Indeed, in our hybrid approach, the Hamiltonian constraint provides a
difference equation in an internal discrete parameter ($v_\theta$ in scheme
A, $v$ in scheme B) which has a strictly positive minimum value, and the
solutions to the Hamiltonian constraint are completely determined by
the data provided on the initial section of such a discrete
parameter. One can say that the solutions follow a
no-boundary prescription, in the sense that they arise in a single
section without the need to impose any particular boundary condition.
This behavior has two important consequences. On the one hand,
this immediately resolves the classical singularity in the
quantum theory at a kinematical level. On the other hand, it
allows one to deal with these solutions by identifying them
with initial data. In this way, we have been able to characterize the
physical Hilbert space in both schemes A and B. Remarkably, as we have
pointed out, this procedure leads to the recovery of
the standard quantum field theory for the inhomogeneities.

\section*{Acknowledgments}

The authors are grateful to D. Brizuela, D. Mart\'in de Blas,
H. Sahlmann, J. Olmedo, T. Pawlowski, E.~Wilson-Ewing,
and specially to J.M.~Velhinho, for useful discussions. This work
was supported by the Spanish MICINN Project
FIS2008-06078-C03-03 and the Consolider-Ingenio 2010
Program CPAN (CSD2007-00042). M. M-B. is supported by CSIC
and the European Social Fund under the grant I3P-BPD2006.

\appendix

\section{Classical metric}
\label{metric-new}

To derive the form of the classical
metric of the linearly polarized Gowdy
$T^3$ model, one can start with its expression in the (field)
parametrization of Ref. \cite{men1}, apply the gauge fixing procedure,
and perform a canonical transformation from the elementary variables
chosen for the homogeneous sector in that parametrization to the
corresponding Ashtekar variables $\{c_{i},p_{i}\}$. A careful
calculation shows that, in our variables, the non-vanishing
components of the induced three-metric are
\begin{align}\label{newmetric}
q_{\theta\theta}&=\frac1{4\pi^2}\left|\frac{p_\sigma
p_\delta}{p_\theta}\right|\exp\left\{\frac{2\pi}
{\sqrt{|p_\theta|}}\bigg(2\frac{c_\delta p_\delta}
{c_\sigma p_\sigma+c_\delta p_\delta}-1\bigg)
\tilde\xi(\theta)\right.\nonumber\\
&\left.-\frac{\pi^2}{|p_\theta|}[\tilde\xi(\theta)]^2
-\frac{8\pi G\gamma}{c_\sigma p_\sigma+c_\delta p_\delta}
\zeta(\theta)\right\},\nonumber\\
q_{\sigma\sigma}&=\frac1{4\pi^2}\left|\frac{p_\theta
p_\delta}{p_\sigma}\right|\exp\left\{-\frac{2\pi}
{\sqrt{|p_\theta|}}\tilde\xi(\theta)\right\},
\nonumber\\q_{\delta\delta}&=\frac1{4\pi^2}
\left|\frac{p_\theta p_\sigma}{p_\delta}
\right|\exp\left\{\frac{2\pi}{\sqrt{|p_\theta|}}
\tilde\xi(\theta)\right\},
\end{align}
where
\begin{align}
\tilde\xi(\theta)&=\frac1{\pi}\sum_{m\neq0}
\sqrt{\frac{G}{|m|}}\big(a_m+a_{-m}^*\big)e^{im\theta},
\nonumber\\
\zeta(\theta)&=i\sum_{m\neq0}\sum_{\tilde m\neq0}
\text{sgn}(m+\tilde m)\frac{\sqrt{|m+\tilde
m||\tilde m|}}{|m|}\nonumber\\&\times \big(a_{-\tilde {m}}-
a^*_{\tilde m}\big)\big(a_{m+\tilde m}+a^*_{-(m+\tilde m)}
\big)e^{im\theta}.\nonumber
\end{align}

Besides, owing to the homogeneity of the shift function
$N^\theta$, required by the gauge fixing,
we can reabsorb it by means of the following redefinition
of the coordinate $\theta$:
\begin{align}
\theta+\int_{t_i}^{t}dt^\prime\,N^\theta(t^\prime)\rightarrow\theta,
\nonumber
\end{align}
where $t_i$ is any initial time. Then, the spacetime metric
becomes
\begin{align}
ds^2=-q_{\theta\theta}\left(\frac{|p_\theta|}{4\pi^2}\right)^2
{N_{_{_{\!\!\!\!\!\!\sim}}\;}}^2dt^2+q_{\theta\theta}d\theta^2
+q_{\sigma\sigma}d\sigma^2
+q_{\delta\delta}d\delta^2.\nonumber
\end{align}
Here, $N_{_{_{\!\!\!\!\!\!\sim}}\;}$ is the densitized lapse function,
which in our gauge is spatially homogenous. The metric of the Bianchi
I spacetime is the result of ignoring the inhomogeneities in Eq.
\eqref{newmetric}, setting $\tilde\xi=0=\zeta$.

\section{Inverse homogeneous volume operator}
\label{invV}

Following the procedures of LQG, the classical expression of
$1/(|p_i|^{1-r})$ in LQC, for $r>0$, is
represented by the regularized operator (see e.g. Ref. \cite{abl})
\begin{align}\label{inv}
\widehat{\left[\!\frac1{|p_i|^{1-r}}\!\right]}\!=\!
\frac{\widehat{\text{sgn}(p_i)}}{8\pi\gamma
l_{\text{Pl}}^2r}\widehat{\left[\frac1{\bar\mu_i}\right]}\!
\big[\hat{\mathcal N}_{-\bar\mu_i}\widehat{|p_i|}^{r}
\hat{\mathcal N}_{\bar\mu_i}-\hat{\mathcal
N}_{\bar\mu_i}\widehat{|p_i|}^{r}\hat{\mathcal N}_{-\bar\mu_i}\big],
\end{align}
where $\widehat{[1/\bar\mu_i]}$ is the quantum counterpart of the
expression given in Eq. \eqref{mubarra} for each of the considered
schemes. The choice of $r$ is arbitrary.

\subsection{Scheme A}
In case A, and for $r=1/2$, one obtains \cite{chio,mmp}
\begin{align}\label{invA}
\widehat{\left[\frac{1}{V}\right]}&=
\otimes_i \widehat{\left[\frac{1}{\sqrt{|p_i|}}\right]}, \quad
\widehat{\left[\frac{1}{\sqrt{|p_i|}}\right]}
|v_i\rangle=b(v_i)|v_i\rangle,\nonumber\\
b(v_i)&=\frac{1}{2(2\pi\gamma l_{\text{Pl}}^2\sqrt{\Delta})^{1/3}}
|v_i|^{\frac1{3}}\left||v_i+1|^{\frac1{3}}-|v_i-1|^{\frac1{3}}\right|.
\end{align}

\subsection{Scheme B}
\label{invVB}

In case B, choosing $r=1/4$ and combining all the powers of
$|p_\theta|$, we obtain the expression
\begin{align}\label{invB}
\widehat{\left[\frac1{|p_\theta|^{\frac1{4}}}\right]}&=
\frac{\widehat{\text{sgn}(p_\theta)}}{2\pi\gamma
l_{\text{Pl}}^2\sqrt{\Delta}}\widehat{\sqrt{|p_\sigma p_\delta|}}
\nonumber\\
&\times\big[\hat{\mathcal N}_{-\bar\mu_\theta}
\widehat{|p_\theta|}^{\frac1{4}}\hat{\mathcal
N}_{\bar\mu_\theta}-\hat{\mathcal
N}_{\bar\mu_\theta}\widehat{|p_\theta|}^{\frac1{4}}
\hat{\mathcal N}_{-\bar\mu_\theta}\big],
\end{align}
and similarly for $\widehat{[1/|p_\sigma|^{1/4}]}$ and
$\widehat{[1/|p_\delta|^{1/4}]}$. Their action on our basis
of states turns out to be
\begin{align}
\widehat{\left[\frac1{|p_i|^{\frac1{4}}}\right]}&|v,
\lambda_\sigma,\lambda_\delta\rangle=
\frac{
b_i(v,\lambda_\sigma,\lambda_\delta)}{(4\pi\gamma
l_{\text{Pl}}^2\sqrt{\Delta})^{\frac1{6}}}|v,
\lambda_\sigma, \lambda_\delta\rangle ,
\end{align}
where $(a=\sigma,\delta)$
\begin{align}\label{btheta}
b_\theta&(v,\lambda_\sigma,\lambda_\delta)=
\sqrt{2|\lambda_\sigma\lambda_\delta|}
\left|\sqrt{|v+1|}-\sqrt{|v-1|}\right|,\nonumber\\
b_a&(v,\lambda_\sigma,\lambda_\delta)=
\sqrt{\left|\frac{v}{\lambda_a}\right|}
\left|\sqrt{|v+1|}-\sqrt{|v-1|}\right|.
\end{align}

The inverse homogeneous volume operator can then be represented
as the regularized operator
\begin{align}
\widehat{\left[\frac{1}{V}\right]}&=
\otimes_i \widehat{\left[\frac{1}{|p_i|^{\frac1{4}}}\right]}^2.
\end{align}

\section{Details of the quantum model for scheme A}
\label{detschemeA}

For completeness in the presentation, we include here some details about
the operator $\widehat{\Theta}_i$ that appears in the expression
\eqref{CGA} of the densitized Hamiltonian constraint operator
$\widehat{\mathcal{C}}_{\text{G}}^\text{A}$ for scheme A.
The definition of $\widehat{\Theta}_i$ in terms of holonomy and
fluxes operators is \cite{mmp,ijmp}:
\begin{eqnarray}
\widehat{\Theta}_i&=&-\frac{i}{4\sqrt{\Delta}}
\widehat{\left[\frac{1}{\sqrt{|p_i|}}
\right]}^{-\frac1{2}}\widehat{\sqrt{|p_i|}} \bigg[(\hat{\mathcal
N}_{2\bar\mu_i}-\hat{\mathcal N}_{-2\bar\mu_i})
\widehat{\text{sgn}(p_i)}\nonumber\\
&+&\widehat{\text{sgn}(p_i)}(\hat{\mathcal
N}_{2\bar\mu_i}-\hat{\mathcal N}_{-2\bar\mu_i})\bigg]
\widehat{\sqrt{|p_i|}}\widehat{\left[\frac{1}{\sqrt{|p_i|}}
\right]}^{-\frac1{2}}.
\end{eqnarray}
Its action on the states $|v_i\rangle$
is given in
Eq. \eqref{acttheta}, where the functions $f_{\pm}(v_i)$ are:
\begin{equation}\label{f}
f_\pm(v_i)=g(v_i\pm2)s_\pm(v_i)g(v_i), \end{equation}
with
\begin{align}
s_\pm(v_i)&=\text{sgn}(v_i\pm2)+\text{sgn}(v_i),\nonumber
\\
g(v_i)&=
\left|\left|1+\frac1{v_i}\right|^{\frac1{3}}
-\left|1-\frac1{v_i}\right| ^{\frac1{3}}
\right|^{-\frac1{2}},\;\text{if }v_i\neq0,\nonumber\\
g(0)&=0.
\end{align}

\section{Details of the quantum model for scheme B}
\label{app:C}

\subsection{The operators $\hat{F}_i$}
\label{operatorF}

Expressing each operator $\hat{F}_i$ in terms of
$\hat{\mathcal N}_{\pm\bar\mu_i}$, a
straightforward
calculation shows that
\begin{align}
&\hat{F}_\theta|v,\lambda_\sigma,\lambda_\delta\rangle=
i\frac{\text{sgn}(\lambda_\sigma\lambda_\delta)}{2}
\sum_{l=+1,-1}l\bigg[\text{sgn}(v)\nonumber\\
&+\text{sgn}\{v+2l\text{sgn}
(\lambda_\sigma\lambda_\delta)\}\bigg]\big|v+2l\text{sgn}
(\lambda_\sigma\lambda_\delta),\lambda_\sigma,
\lambda_\delta\big\rangle,
\end{align}
\begin{align}\label{Fsigma}
&\hat{F}_\sigma|v,\lambda_\sigma,\lambda_\delta\rangle=i
\frac{\text{sgn}(\lambda_\sigma)}{2}
\sum_{l=+1,-1}l\bigg[\text{sgn}\{|v|+2l\text{sgn}
(\lambda_\sigma)\}\nonumber\\
&+1\bigg]\bigg|v+2l\text{sgn}
(v\lambda_\sigma),\lambda_\sigma+2l
\left|\frac{\lambda_\sigma}{v}\right|,
\lambda_\delta\bigg\rangle.
\end{align}
The action of $\hat{F}_\delta$ is similar to that of $\hat{F}_\sigma$,
interchanging the roles of $\lambda_\sigma$ and $\lambda_\delta$.

\subsection{The operator $\widehat{\mathcal{C}}_{\text{G}}^\text{B}$}
\label{operatorCB}

The action of the constraint operator on the states
$|v,\lambda_\sigma,\lambda_\delta\rangle\otimes|\mathfrak{
N}\rangle$ of our basis, with $v$, $\lambda_\sigma$,
and $\lambda_\delta$ being all positive, is
the following:
\begin{align}\label{actionCB}
&\widehat{\mathcal{C}}_{\text{G}}^\text{B}|v,
\lambda_\sigma,\lambda_\delta\rangle\otimes|\mathfrak{
N}\rangle=\frac{(\pi
l_{\text{Pl}}^2)^2}{4}\nonumber\\&\times
\bigg\{x_-(v)|v-4,\lambda_\sigma,\lambda_\delta\rangle_-
-x^-_0(v)|v,\lambda_\sigma,\lambda_\delta\rangle_{0^-}
\nonumber\\&-x^+_0(v)|v,\lambda_\sigma,
\lambda_\delta\rangle_{0^+}+x_+(v)|v+4,\lambda_\sigma,
\lambda_\delta\rangle_+\nonumber\\
&+\frac{32}{\beta}\frac{v^2}{\lambda_\sigma^2
\lambda_\delta^2}\widehat{H}_0^\xi|v,\lambda_\sigma,
\lambda_\delta\rangle-
\beta b_\theta^2(v,\lambda_\sigma,\lambda_\delta)
\widehat{H}_\text{int}^\xi\nonumber\\
&\times\bigg[
b_\theta^2(v-4,\lambda_\sigma,\lambda_\delta)
\frac{v-4}{v}x_-(v)|v-4,\lambda_\sigma,
\lambda_\delta\rangle_-'\nonumber\\
&-b_\theta^2(v,\lambda_\sigma,\lambda_\delta)\big[x^-_0(v)|v,
\lambda_\sigma,\lambda_\delta\rangle_{0^-}'+
x^+_0(v)|v,\lambda_\sigma,\lambda_\delta\rangle_{0^+}'\big]
\nonumber\\&+b_\theta^2(v+4,\lambda_\sigma,\lambda_\delta)
\frac{v+4}{v}x_+(v)|v+4,\lambda_\sigma,
\lambda_\delta\rangle_+'\bigg]\bigg\}\otimes|\mathfrak{N}
\rangle,
\end{align}
where
\begin{align}\label{beta}
\beta=\left(\frac{l_{\text{Pl}}}
{4\pi\gamma\sqrt{\Delta}}\right)^{2/3}
\end{align}
and we have introduced the following notation:
\begin{align}\label{state1}
&|v\pm4,\lambda_\sigma,\lambda_\delta\rangle_\pm
\nonumber\\
&=\bigg|v\pm4,\lambda_\sigma,\frac{v\pm4}
{v\pm2}\lambda_\delta\bigg\rangle+
\bigg|v\pm4,\lambda_\sigma,\frac{v\pm2}
{v}\lambda_\delta\bigg\rangle\nonumber\\
&+\bigg|v\pm4,\frac{v\pm4}
{v\pm2}\lambda_\sigma,\frac{v\pm2}
{v}\lambda_\delta\bigg\rangle+
\bigg|v\pm4,\frac{v\pm4}
{v\pm2}\lambda_\sigma,\lambda_\delta
\bigg\rangle\nonumber\\&+\bigg|v\pm4,\frac{v\pm2}
{v}\lambda_\sigma,\frac{v\pm4}
{v\pm2}\lambda_\delta\bigg\rangle+
\bigg|v\pm4,\frac{v\pm2}{v}\lambda_\sigma,\lambda_\delta
\bigg\rangle,
\end{align}
\begin{align}
& |v,\lambda_\sigma,\lambda_\delta\rangle_{0^\pm}
\nonumber\\&=\bigg|v,\lambda_\sigma,\frac{v}
{v\pm2}\lambda_\delta\bigg\rangle+
\bigg|v,\lambda_\sigma,\frac{v\pm2}
{v}\lambda_\delta\bigg\rangle\nonumber\\
&+\bigg|v,\frac{v}{v\pm2}
\lambda_\sigma,\lambda_\delta\bigg\rangle+
\bigg|v,\frac{v}{v\pm2}\lambda_\sigma,\frac{v\pm2}
{v}\lambda_\delta\bigg\rangle\nonumber\\
&+\bigg|v,\frac{v\pm2}{v}\lambda_\sigma,\lambda_\delta
\bigg\rangle+\bigg|v,\frac{v\pm2}
{v}\lambda_\sigma,\frac{v}
{v\pm2}\lambda_\delta\bigg\rangle,
\end{align}
\begin{align}
&
|v\pm4,\lambda_\sigma,\lambda_\delta\rangle_\pm'\nonumber\\
&=\bigg|v\pm4,\lambda_\sigma,\frac{v\pm4}
{v}\lambda_\delta\bigg\rangle+\bigg|v\pm4,\frac{v\pm4}
{v\pm2}\lambda_\sigma,\frac{v\pm2}
{v}\lambda_\delta\bigg\rangle\nonumber\\
&+\bigg|v\pm4,\frac{v\pm4}{v}\lambda_\sigma,\lambda_\delta
\bigg\rangle+\bigg|v\pm4,\frac{v\pm2}
{v}\lambda_\sigma,\frac{v\pm4}
{v\pm2}\lambda_\delta\bigg\rangle,
\end{align}
\begin{align}\label{state4}
&|v,\lambda_\sigma,\lambda_\delta\rangle_{0^\pm}'=
2|v,\lambda_\sigma,
\lambda_\delta\rangle+
\bigg|v,\frac{v\pm2}{v}\lambda_\sigma,\frac{v}
{v\pm2}\lambda_\delta\bigg\rangle\nonumber\\&+
\bigg|v,\frac{v}{v\pm2}\lambda_\sigma,\frac{v\pm2}
{v}\lambda_\delta\bigg\rangle,
\end{align}
and
\begin{align}
x_-(v)&=2\sqrt{v}(v-2)\sqrt{v-4}[1+\text{sgn}(v-4)],
\label{coefficient1}\\
x_+(v)&=x_-(v+4), \label{coefficient2}\\
x^-_0(v)&=2(v-2)v[1+\text{sgn}(v-2)],
\label{coefficient3}\\ x^+_0(v)&=x^-_0(v+2)
\label{coefficient4}.
\end{align}

\subsection{Support of the anisotropies}
\label{proof}

We want to prove that the set $\mathcal W_\varepsilon$ is dense in
$\mathbb{R}^+$. For this, we will show that its
subset $U_\varepsilon$,
defined as
\begin{equation}
\mathcal W_\varepsilon\supset U_\varepsilon=
\left\{\frac{\varepsilon+4m}{\varepsilon+4n};\,
m,n\in\mathbb{N}\right\},
\end{equation}
is already dense in the positive real line.

Let $a$ and $b$ be any two positive real numbers, such that $b-a>0$.
Besides, we define the set
\begin{equation}
V_\varepsilon=\left\{\frac{\varepsilon}{4}+n;\, n\in\mathbb{N}\right\}.
\end{equation}
Then there always exists a number
$s=\varepsilon/4+n_1\in V_\varepsilon$ such that $1<s(b-a)$,
or equivalently
\begin{align}\label{a}
 sa+1<sb.
\end{align}
Let us denote by $t=\varepsilon/4+m_1$
the largest number in $V_\varepsilon$ which is smaller or equal than $sa+1$,
that is
\begin{equation}\label{b}
 sa< \frac{\varepsilon}{4}+m_1 \leq sa+1.
\end{equation}
Equations \eqref{a} and \eqref{b} imply that $sa<\varepsilon/4+m_1<sb$.
These inequalities can be written equivalently as
\begin{align}
 a<\frac{\varepsilon+4m_1}{\varepsilon+4n_1}<b.
\end{align}
In conclusion, given any two positive numbers $a$ and $b$ with $a<b$,
there always exists a number $u\in U_\varepsilon$ such that $a<u<b$.
As a consequence $U_\varepsilon$, and $\mathcal
W_\varepsilon$ a fortiori, are dense in the positive real axis, as we
wanted to prove.


\begin{thebibliography}{99}

\bibitem{lqg} T. Thiemann, {\it{Modern Canonical Quantum General
Relativity}} (Cambridge University Press, Cambridge, England, 2007);
C. Rovelli, {\it{Quantum Gravity}} (Cambridge University Press,
Cambridge, England, 2004); A. Ashtekar and J. Lewandowski, Classical
Quantum Gravity {\bf 21}, R53 (2004).

\bibitem{lqc} M. Bojowald, Living Rev. Relativity {\bf 11}, 4 (2008).

\bibitem{abl} A. Ashtekar, M. Bojowald, and J. Lewandowski, Adv. Theor.
Math. Phys. {\bf 7}, 233 (2003).

\bibitem{aps3} A. Ashtekar, T. Paw{l}owski, and P. Singh, Phys. Rev. D
{\bf 74}, 084003 (2006).

\bibitem{mmo} M. Mart\'{\i}n-Benito, G.A. Mena Marug\'{a}n, and J.
Olmedo, Phys. Rev. D {\bf80}, 104015 (2009).

\bibitem{iso1} A. Ashtekar, T. Pawlowski, P. Singh, and K. Vandersloot,
Phys. Rev. D {\bf 75}, 024035 (2007).

\bibitem{iso2} L. Szulc, W. Kaminski, and J. Lewandowski, Classical
Quantum Gravity {\bf 24}, 2621 (2007); K. Vandersloot, Phys. Rev. D
{\bf 75}, 023523 (2007).

\bibitem{iso3}E. Bentivegna and T. Pawlowski, Phys. Rev. D {\bf 77},
124025 (2008).

\bibitem{chio} D.W. Chiou, Phys. Rev. D {\bf 75}, 024029 (2007).

\bibitem{chi2} D.W. Chiou, Phys. Rev. D {\bf 76}, 124037 (2007).

\bibitem{mmp} M. Mart\'{\i}n-Benito, G.A. Mena Marug\'{a}n, and
T. Paw\-lowski, Phys. Rev. D {\bf78}, 064008 (2008).

\bibitem{awe1} A. Ashtekar and E. Wilson-Ewing, Phys. Rev. D
{\bf 79}, 083535 (2009).

\bibitem{awe2} A. Ashtekar and E. Wilson-Ewing, Phys. Rev. D
{\bf 80}, 123532 (2009).

\bibitem{gowd} R.H. Gowdy, Ann. Phys. {\bf83}, 203 (1974).

\bibitem{mon} V. Moncrief, Phys. Rev. D {\bf23}, 312 (1981).

\bibitem{ise} J. Isenberg and V. Moncrief, Ann. Phys. {\bf 199},
84 (1990).

\bibitem{qGow} See, e.g., C.W. Misner, Phys. Rev. D {\bf8},
3271 (1973); B.K. Berger, Ann. Phys. {\bf83}, 458 (1974);
Phys. Rev. D {\bf 11}, 2770 (1975); Ann. Phys. {\bf 156}, 155 (1984);
G.A. Mena Marug\'{a}n, Phys. Rev. D {\bf56}, 908 (1997); M. Pierri,
Int. J. Mod. Phys. D {\bf11}, 135 (2002).

\bibitem{men1} A. Corichi, J. Cortez, and G.A. Mena Marug\'{a}n,
Phys. Rev. D {\bf73}, 041502 (2006); A. Corichi, J. Cortez, and
G.A. Mena Marug\'{a}n, Phys. Rev. D {\bf73}, 084020(R) (2006).

\bibitem{men3} A. Corichi, J. Cortez, G.A. Mena Marug\'{a}n, and
J. M. Velhinho, Classical Quantum Gravity {\bf23}, 6301 (2006);
J. Cortez, G.A. Mena Marug\'{a}n, and J. M. Velhinho, Phys. Rev. D
{\bf75}, 084027 (2007).

\bibitem{let} M. Mart\'{\i}n-Benito, L.J. Garay, and G.A. Mena
Marug\'{a}n, Phys. Rev. D {\bf78}, 083516 (2008).

\bibitem{ijmp} G.A. Mena Marug\'{a}n and  M. Mart\'{\i}n-Benito,
Int. J. Mod. Phys. A {\bf24}, 2820 (2009).

\bibitem{man} G.A. Mena Marug\'{a}n and M. Montejo, Phys. Rev. D
{\bf58}, 104017 (1998).

\bibitem{note} Here, we call physical area that measured by
the area operator defined on the kinematical Hilbert space, to
distinguish it from the fiducial area.



\bibitem{note1} There is a discrepancy in signs between this
representation and that of Ref. \cite{awe1} because, in
the latter, $\gamma$ is considered to change
sign under internal parity transformations, namely
$\gamma=|\gamma|\text{sgn}(v)$, while we treat $\gamma$ just
as a positive free parameter, which therefore is unaffected
by transformations of the dynamical variables.

\bibitem{note2} We understand {\sl non-densitized} Hamiltonian
constraint to refer to the scalar constraint with the same
densitization as in LQG.

\bibitem{note3} This difference is due to the factor $1/(16\pi G)$
accompanying the classical action, which is not included in the
definition of the constraint in this paper.

\bibitem{group} D. Marolf, \texttt{arXiv:gr-qc/9508015}; Classical
Quantum Gravity {\bf 12}, 1199 (1995); {\bf 12}, 1441
(1995); {\bf 12}, 2469, (1995).

\bibitem{red} A. D. Rendall, Classical Quantum Gravity {\bf 10}, 2261
(1993); \texttt{arXiv:gr-qc/9403001}.

\bibitem{eff} D. Brizuela, G.A. Mena Marug\'an, and T. Pawlowski,
Classical Quantum Gravity {\bf 27}, 52001 (2010).

\bibitem{note4} Nonetheless, we note that
 $\widehat{\sqrt{V}}(\hat{F}_i-\hat{F}_j)\widehat{\sqrt{V}}$
is still a constant of motion. We thank T. Pawlwoski for pointing
out this fact.

\bibitem{mmw} M. Mart\'{\i}n-Benito, G.A. Mena Marug\'{a}n, and
E. Wilson-Ewing, Hybrid Quantization: From Bianchi I to
the Gowdy Model ({\it{in preparation}}).

\bibitem{mpw} G.A. Mena Marug\'{a}n, T. Pawlowski, and
E. Wilson-Ewing ({\it{in preparation}}).


\end{thebibliography}
\end{document}